\documentclass[conference]{IEEEtran}
\IEEEoverridecommandlockouts
\usepackage{flushend}
\usepackage{float}
\usepackage{cite}
\usepackage{amsmath,amssymb,amsfonts}
\usepackage{algorithm}
\usepackage{algcompatible}
\newcommand{\argmin}{\mathrm{argmin}}
\newcommand{\R}{\mathbb{R}}

\usepackage{amsthm}
\newtheorem{proposition}{Proposition}

\usepackage{subfigure}
\usepackage{bm}
\usepackage{multirow}
\usepackage{graphicx}
\usepackage{textcomp}
\def\BibTeX{{\rm B\kern-.05em{\sc i\kern-.025em b}\kern-.08em
    T\kern-.1667em\lower.7ex\hbox{E}\kern-.125emX}}

\newcommand{\tabincell}[2]{\begin{tabular}{@{}#1@{}}#2\end{tabular}}

\begin{document}

\title{RobustScaler: QoS-Aware Autoscaling\\ for Complex Workloads
}

\author{\IEEEauthorblockN{Huajie Qian, Qingsong Wen, Liang Sun, Jing Gu, Qiulin Niu, Zhimin Tang}
\IEEEauthorblockA{\textit{DAMO Academy, Alibaba Group} \\
Bellevue, WA, USA \\
\{h.qian, qingsong.wen, liang.sun, zibai.gj, qiulin.nql, zhimin.tangzm\}@alibaba-inc.com}
}

\maketitle

\begin{abstract}
Autoscaling is a critical component for efficient resource utilization with satisfactory quality of service (QoS) in cloud computing. This paper investigates proactive autoscaling for widely-used scaling-per-query applications where scaling is required for each query, such as container registry and function-as-a-service (FaaS). In these scenarios, the workload often exhibits high uncertainty with complex temporal patterns like periodicity, noises and outliers. Conservative strategies that scale out unnecessarily many instances lead to high resource costs whereas aggressive strategies may result in poor QoS. We present RobustScaler to achieve superior trade-off between cost and QoS. Specifically, we design a novel autoscaling framework based on non-homogeneous Poisson processes (NHPP) modeling and stochastically constrained optimization. Furthermore, we develop a specialized alternating direction method of multipliers (ADMM) to efficiently train the NHPP model, and rigorously prove the QoS guarantees delivered by our optimization-based proactive strategies. Extensive experiments show that RobustScaler outperforms common baseline autoscaling strategies in various real-world traces, with large margins for complex workload patterns.
\end{abstract}


\section{Introduction}
Autoscaling has been a fundamental tool in elastic cloud services that dynamically adds (scales out) or deletes (scales in) computing resources, such as instances, CPU and memory, to closely match the ever-changing computing demand~\cite{herbst2013elasticity,qu2018auto}. Real-world workload can change drastically over time, and reactive autoscaling that conducts scaling actions only after the workload changes often leads to underprovision during increasing traffic and in turn degradation of quality of service (QoS). Therefore, proactive autoscaling is becoming increasingly popular~\cite{taft2018pstore,rebjock2020simple,rzadca2020autopilot,tsoumakos2013automated} in that it scales out beforehand when a demand increase is expected so that resource shortfalls can be avoided and satisfactory QoS is consistently maintained.

In this paper, we consider proactive autoscaling in the \textit{scaling-per-query} scenario of on-demand cloud service: When a query comes in, it will be served by one of the available idle instances (warm start) or a new instance will be initiated for that query (cold start); meanwhile, each instance is terminated after processing one query and is not reused/shared by other queries. 
The scaling-per-query scenario is common in cloud computing~\cite{mahmoudi2020performance, wong2021threat}: One example is the container registry~\cite{wang2021faasnet, anwar2018improving} for storing, managing, and securing custom container images; another example can be found in Continuous Integration (CI) and Continuous Delivery (CD)~\cite{hakli2018towards} to avoid user’s account/data being hijacked/spoofed.
Note that scaling-per-query is in contrast to other cloud services where an instance keeps running and processing new queries until it is terminated for other reasons (e.g., scaled in when traffic goes down). 
A purely reactive scaling mechanism in the scaling-per-query scenario would simply create an instance for every new query and terminate the instance after processing the query. The main drawback of this mechanism is the cold-start delay, i.e., the startup time of an instance is not negligible compared to the actual processing time of the query, making the total response time unnecessarily long. One way to reduce the cold-start delay is to maintain a pool of running instances in anticipation of future queries so that processing can start immediately upon arrival of a query, however, this can incur a considerable overhead cost since the instances may stay idle without processing any query, especially when the pool size is large, and it is not clear what pool size is needed to achieve a certain QoS. 

In practice, proactive autoscaling also encounters several challenges. One challenge lies in capturing the complex and variable periodic patterns in the workload. The workloads of database and cloud computing often exhibit notable periodic patterns~\cite{yan2021workload,higginson2020database,atikoglu2012workload,cortez2017resource,calzarossa2016workload,SPAR08}. By identifying and utilizing the periodicity, we can perform effective autoscaling of resources in these scenarios to save a significant amount of resources. However, real-world workload may come with lots of missing data, anomalies, large noises that, on one hand, obscure many structural patterns such as periodicity that are crucial for prediction, and on the other hand requires strong robustness from the workload prediction algorithm. 
We leverage robust decomposition~\cite{WenRobustPeriod20,wen2020fastrobustSTL} to extract periodic patterns from the queries-per-second (QPS) time series for query arrival modeling and forecasting, even in the presence of a considerable amount of noises, missing data and anomalies.
Another challenge concerns the control of the trade-off between resource saving and QoS improvement. As more resources are allocated, the QoS gets improved, and vice versa. For scaling-per-query, the more instances are maintained and the earlier they are created, the less cold starts will occur. Proper control of this trade-off becomes more challenging for scaling-per-query due to its per-query scaling dynamics. For example, the lifecycle (hence the cost) of an instance also depends on the arrival and the execution time of the query, and whether cold start occurs is related to the temporal order between the arrival time of the query and the time the instance finishes startup which is not known beforehand when creating the instance. Therefore, even evaluating the cost and QoS (e.g., cold-start delay) requires knowledge of the query arrival dynamics. Moreover, instances created before will be consumed by incoming queries and therefore the number of available instances changes after every query arrival, making scaling-per-query a sequential decision making problem. All these necessitate the modeling of the query arrival process for which we develop a non-homogeneous Poisson process (NHPP) framework with novel periodicity regularization that can flexibly approximate variable periodicity patterns. Moreover, we formally investigate the trade-off through the lens of stochastically constrained optimization~\cite{birge2011introduction} from which scaling decisions that respect either QoS constraints or cost constraints can be efficiently computed.

The main goal of this study is a proactive autoscaling framework that can properly create instances before the queries arrive to guarantee a certain QoS with a low overhead cost. To the best of our knowledge, no previous work has investigated proactive autoscaling to reduce cold-start delay in the scaling-per-query scenario. To this end, we design a novel autoscaling framework called RobustScaler which can generate robust scaling decisions that optimally balance the trade-off between cost and QoS and are robust to noise, missing data and anomalies. In summary, our contributions are: 1) develop the first NHPP framework for demand modeling in autoscaling to capture both periodicity and stochasticity of query arrivals, along with an efficient alternating direction method of multipliers (ADMM) training algorithm; 2) propose a stochastically constrained optimization formulation to compute Pareto-optimal scaling decisions in terms of cost and QoS; 3) design a sequential scaling scheme based on the stochastically constrained optimization that enjoys provable probabilistic QoS guarantees; 4) conduct an extensive set of experiments that demonstrate the superiority of RobustScaler compared to heuristic strategies.

The rest of the paper is organized as follows. Section \ref{sec:related work} reviews the related literature. Section \ref{sec:problem statement} introduces the setting and challenges for scaling-per-query. Section \ref{sec:framework} describes the proposed autoscaling framework. Section \ref{sec:point process modeling} then presents our NHPP arrival model with periodicity regularization, whereas Section \ref{sec:scaling decision} details the QoS-cost trade-off in scaling decision making, the constrained optimization formulations, and the consequent sequential scaling schemes. Section \ref{sec:experiments} presents experimental results including a comparison of the proposal with two heuristic autoscaling methods. Section \ref{sec:conclusion} concludes the paper.


\section{Related Work}\label{sec:related work}

Autoscaling is an active research area in various types of cloud systems, such as database~\cite{lolos2017adaptive,jindal2019peregrine,taft2018pstore}, microservices~\cite{yu2020microscaler,abdullah2020burst,bauer2019chamulteon}, stream processing~\cite{mei2020turbine,wang2019elasticutor,singh2020auto}, and web applications~\cite{jiang2013optimal,aslanpour2017auto}. Recent surveys include~\cite{lorido2014review,al2017elasticity,qu2018auto,chen2018survey,barnawi2020views}. These autoscaling systems have been successfully applied in their respective types of systems, but cannot handle the scaling-per-query scenario considered in this paper where the key difficulty is to mitigate cold start for future queries. Similar cold start issues also appear in serverless FaaS platforms, and are typically mitigated by either maintaining a pool of pre-warmed instances ~\cite{mohan2019agile,lin2019mitigating} or reusing warm instances after they finish current function invocations~\cite{wang2018peeking}. However, unlike the serverless environment, instances are not reused in our case and hence the scaling dynamics are different.


Based on the scaling timing, autoscaling methods can be categorised into proactive and reactive ones~\cite{qu2018auto}. Note that the full potential of proactive autoscaling can be utilized when the workload has predictable and periodic/cyclic patterns. Fortunately, many real-world workloads exhibit such patterns~\cite{yan2021workload,higginson2020database,atikoglu2012workload,cortez2017resource,calzarossa2016workload,SPAR08}. However, in practice we often face highly noisy and variable periodic patterns with missing data and outliers~\cite{taft2018pstore,WenRobustPeriod20}, which calls for robust workload prediction. 
For example, P-Store~\cite{taft2018pstore} is a predictive autoscaling scheme for database systems under workload with a diurnal pattern, but it does not automatically capture other periodicity patterns. Meanwhile, P-store does not characterize uncertainties in workload forecasting.
Periodic workload patterns are also utilized in the Turbine of Facebook~\cite{mei2020turbine} for stream processing systems, where it forecasts future load patterns in a heuristic way without carefully considering pattern variations or outliers. In contrast, we leverage robust decomposition~\cite{WenRobustPeriod20,wen2020fastrobustSTL} of time series in workload forecasting to deal with complex variable periodic patterns under challenging noise and outliers.

To properly characterize uncertain query arrivals in the workload, a conventional modeling tool is Poisson processes~\cite{calzarossa1985characterization,mahmoudi2020performance} that is analytically tractable with elegant statistical properties and accurate enough when the traffic consists of many independent users with weak temporal dependency \cite{pitchumani2015realistic}. Otherwise, Markovian arrival processes \cite{pacheco2011markovian} and hierarchical bundling models \cite{juan2014beyond} have been developed to capture bursts and temporal dependency. See \cite{calzarossa2016workload} for a comprehensive survey. However, in these approaches periodicity patterns are either not considered or assumed not to vary over time, whereas we adopt NHPP with a novel periodicity regularization to flexibly approximate variable periodicity patterns.


Existing autoscalers generate scaling decisions based on control theory~\cite{roy2011efficient,jamshidi2016managing,ullah2018control}, reinforcement learning~\cite{rossi2019horizontal,gari2020reinforcement}, queuing theory~\cite{moreno2019efficient,gandhi2018model}, and rule-based methods~\cite{taherizadeh2019dynamic}.
Most of these methods either make scaling decisions based on a mean demand estimate without considering uncertainty, or handle uncertainty in a heuristic way. 
For example, model predictive control is adopted for predictive autoscaling in~\cite{roy2011efficient} which utilizes ARMA model for workload forecasting and look-ahead controller for resource allocation without considering uncertainty. The autoscaling scheme RobustT2Scale~\cite{jamshidi2016managing} integrates a fuzzy controller with an online learning mechanism that can cope with certain uncertainties but is not general enough to handle variable periodic patterns. Instead we directly incorporate workload uncertainty into scaling decisions via stochastic constraints that are expressive of QoS requirements to derive robust scaling decisions.





\section{The Autoscaling Problem}\label{sec:problem statement}
\begin{figure}
    \centering
    \includegraphics[width=0.99\linewidth]{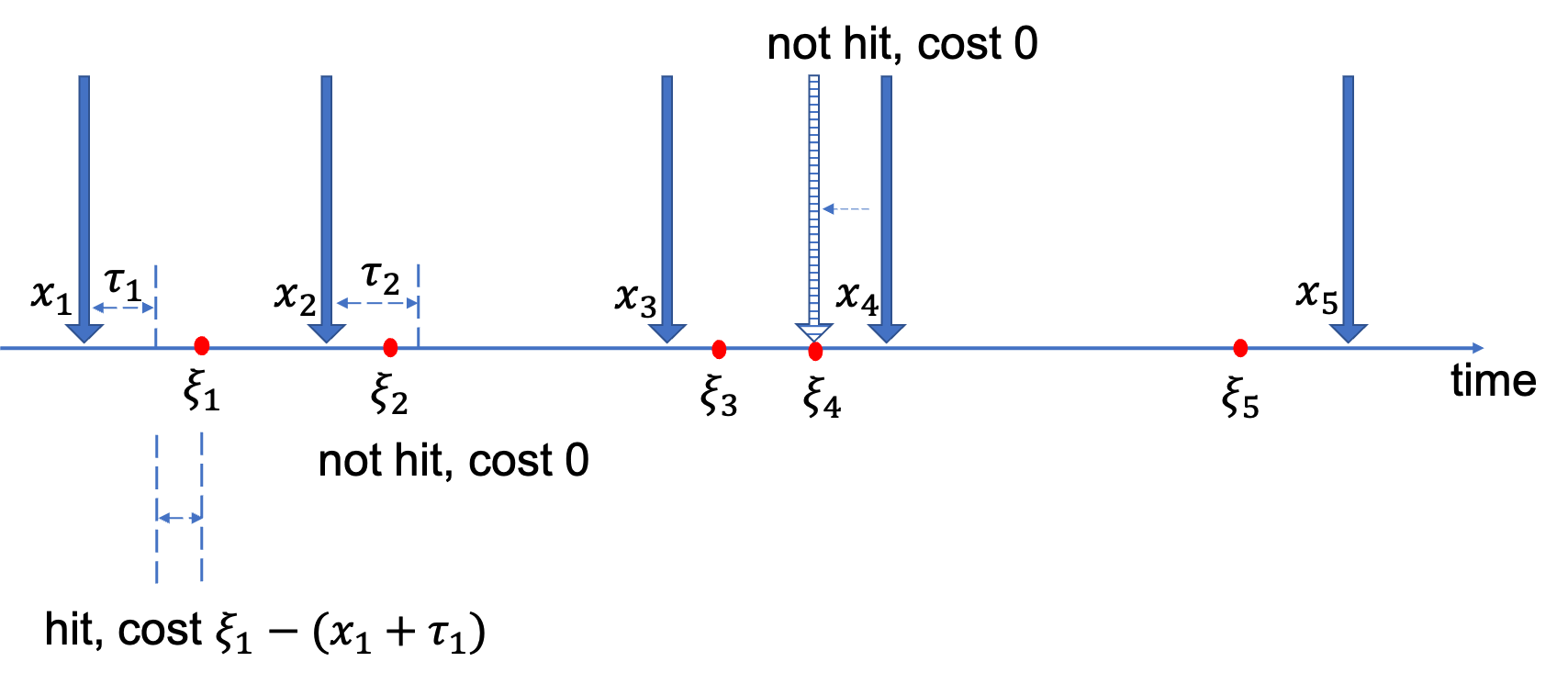}
    \vspace{-8mm}
    \caption{The scaling process. Red dots denote query arrivals and blue arrows denote instance creation.} 
    \label{fig:arrival process illustration}
\end{figure}

Formally, let $\Delta t$ be a fixed time step, and $\{Q_t: t=1,\ldots,T\}$ be a sequence of historical query counts from an application deployment within each time interval of length $\Delta t$, that is, $\frac{Q_t}{\Delta t}$ is the QPS. 
Suppose the current time is $0$, and let
$0<\xi_1<\xi_2<\cdots<\xi_i<\cdots$
be an ascending sequence of random arrival times of the queries, where $\xi_i$ is the time of arrival of the $i$-th upcoming query, let
$0<x_1< x_2 <\cdots<x_i<\cdots$
be an ascending sequence of deterministic instance creation times (the $i$-th instance to be used to process the $i$-th query). Let
$s_1,\ s_2,\ldots,\ s_i,\ldots$ be a sequence of i.i.d. random processing times of the queries, and $\tau_1,\ \tau_2,\ldots,\ \tau_i,\ldots$ be a sequence of i.i.d. random pending/startup times of the instances. Therefore the $i$-th instance shall get ready for processing queries at time $x_i+\tau_i$. Let
$\mu_{s}:=E[s_i],\ \mu_{\tau}=E[\tau_i]$
be the expected values of each $s_i, \tau_i$ respectively. To illustrate the dynamics of the query arrivals, their interplay with scaling actions, and the associated resource cost, Figure \ref{fig:arrival process illustration} shows a sample of the query arrival process.
If the arrival time $\xi_i$ of the $i$-th query is later than the time $x_i+\tau_i$ when the $i$-th instance gets ready (e.g., the first query in Figure \ref{fig:arrival process illustration}), the query is said to be ``hit'' and can be processed right away, but the instance stays idle for a time span of $\xi_i-(x_i+\tau_i)$ which incurs an unnecessary cost. If the arrival time is earlier than the ready time (e.g., the second query), the query is not hit and needs to wait until the instance gets ready to start processing. Otherwise if the $i$-th instance is not even created before $\xi_i$ (e.g., the fourth query), it will be immediately created to handle the query and the originally scheduled creation at time $x_i$ is canceled. Finally, once the instance finishes processing the query, it gets deleted immediately. These dynamics are formalized in Algorithm \ref{algo:dynamics}.

\begin{algorithm}[t]
\caption{Scaling dynamics}
\label{algo:dynamics}
\begin{algorithmic}[1]
\STATEx Given the instance creation times $x_1,x_2,\ldots$, 
\FOR{$i=1,2,\ldots$ -th query}
\IF{$x_i+\tau_i\leq \xi_i$}
\STATE the query is processed immediately
\ELSIF{$x_i\leq \xi_i<x_i+\tau_i$}
\STATE the query waits until $x_i+\tau_i$ and then processing starts
\ELSE
\STATE an instance is created to process the query
\ENDIF
\ENDFOR
\end{algorithmic}
\end{algorithm}

Our framework aims to recommend the scaling action sequence $0<x_1< x_2 <\cdots<x_i<\cdots$ based on historical data so that
\begin{itemize}
    \item (QoS) the response time of a query, i.e., the time between its arrival and the completion of its processing, is short,
    \item (cost) the idling time of a instance, i.e., the period between the time it gets fully started and the time it starts to process a query, is short.
\end{itemize}
Note that these two metrics compete against each other as illustrated before in Figure \ref{fig:arrival process illustration}, and balancing their trade-off serves as a fundamental challenge in making scaling decisions.

\section{Framework of RobustScaler}\label{sec:framework}
\begin{figure}[h]
    \centering
    \includegraphics[width=0.99\linewidth]{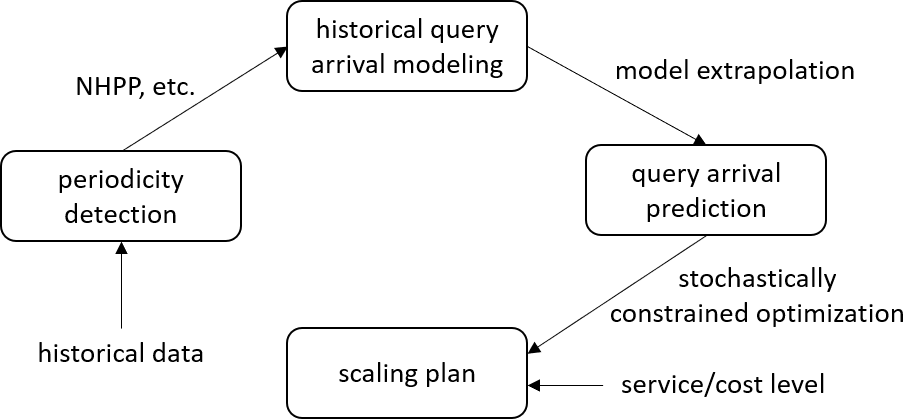}
    \caption{Framework of the proposed RobustScaler method.} 
    \label{fig:framework}
\end{figure}
In this section we introduce the framework of the proposed RobustScaler algorithm. Specifically, it consists of four main components, including periodicity detection, historical query arrival modeling, query arrival prediction, and scaling plan, as illustrated in Figure \ref{fig:framework}. In the following we introduce each component in detail. 

The first \emph{periodicity detection} module takes in historical query data and discovers any present cyclic patterns by leveraging robust periodicity detection~\cite{WenRobustPeriod20}. Depending on the size of fluctuation of the traffic and the time resolution of the data, periodicity patterns can be obscured by the inherent randomness of the traffic and thus can not be easily detected. Therefore, to reduce random effects and reveal potentially hidden periodicity patterns, we perform a time aggregation by averaging the QPS series in each time window of a fixed size and then perform periodicity detection on the aggregated series.


The second \emph{historical query arrival modeling} module is the key modeling step that learns the stochasticity of the query arrival process and makes it possible 1) to make fine-grained query-wise scaling decisions and 2) to make robust scaling decisions that are guaranteed to attain a certain QoS level under traffic uncertainty. In particular, a regularized NHPP with periodicity penalty is used, a more detailed exposition for which shall be given in Section \ref{sec:point process modeling}. Note that this step on its own is a workload modeling technique that is generally applicable to other settings, although the focus of this work is on scaling-per-query.

The third \emph{query arrival prediction} module extrapolates the trained point process model to infer the dynamics of the incoming traffic in the near future. In the case of NHPP, the estimated historical intensity is extrapolated to predict future intensity of the arrival process.

The last \emph{scaling plan} module utilizes the predicted traffic dynamics to create scaling plans, i.e., a list of scaling in/out actions to be performed in the near future, through a stochastically constrained optimization formulation that characterizes a trade-off between resource cost and QoS. User-specified QoS or cost levels can be directly fed into the formulation to obtain scaling decisions that satisfy the stipulated QoS/cost requirements. Section \ref{sec:scaling decision} discusses in detail the stochastically constrained formulation and our sequential scaling scheme.

\section{Query Arrival Modeling}\label{sec:point process modeling}
\vspace{-1mm}
This section explains the details of our second module that models the query traffic as a point process~\cite{mahmoudi2020performance}. Specifically, we consider modeling query arrivals as an NHPP. Suppose that within each time step $\Delta t$ the Poisson intensity is a constant $\lambda_t:=\exp(r_t)$, where an exponential transform is adopted to ensure the positiveness of the intensity, so that the query count during the $t$-th time interval has a Poisson distribution with rate $\exp(r_t)\Delta t$. Then the likelihood of observing $Q_t$ queries within the $t$-th time interval is ${\exp(-\exp(r_t)\Delta t)(\exp(r_t)\Delta t)^{Q_t}}/{Q_t!}$ and the negative log-likelihood of query count series $\{Q_t: t=1,\ldots,T\}$ is
$lkh(\mathbf{r}) \propto -\mathbf{Q}^T\mathbf{r} + \Delta t\cdot \mathbf{1}^T\exp(\mathbf{r})$,
where $\mathbf{Q}=(Q_1,\ldots,Q_T)^T$, $\mathbf{r}=(r_1,\ldots,r_T)^T$, $\exp(\mathbf{r})$ is the element-wise exponential of $\mathbf{r}$, and $\mathbf{1}$ is the all-one vector of length $T$.
With the likelihood loss alone, the learned intensities will be identical to the QPS ${Q_t}/{\Delta t}$ for each $t$ which are susceptible to noises and outliers and cannot handle periodic patterns. Therefore, we impose an $L_1$ penalty on the second order difference to regularize the intensity curve, and further introduce periodicity by imposing an additional periodicity regularization term if periodicity patterns are detected, arriving at a regularized loss
\begin{equation}\label{NHPP formulation w/ periodicity}
\min_{\mathbf{r}} \ \ -\mathbf{Q}^T\mathbf{r} + \Delta t\cdot \mathbf{1}^T e^{\mathbf{r}} + \beta_1 \Vert D^2\mathbf{r} \Vert_1 + \frac{\beta_2}{2} \Vert D_L\mathbf{r} \Vert_2^2\vspace{-2mm}
\end{equation}
where the matrix $D^2$
is the second order differential matrix to capture the smoothness between every set of three consecutive points~\cite{kim2009ell_1,hodrick1997postwar}, $\beta_1\geq 0$ is the smoothness regularization parameter, $L$ is the period length obtained from periodicity detection, the matrix
$D_L=\left[I_{T-L},\ \mathbf{0}_{(T-L)\times L}\right] - \left[\mathbf{0}_{(T-L)\times L},\ I_{T-L}\right] \in \R^{(T-L) \times T}$
is the $L$-step forward differential matrix to capture the smoothness across period length $L$, and $\beta_2$ is the periodicity regularization parameter. Additionally, the time step $\Delta t$ for the QPS series also affects the accuracy of the arrival model. Ideally, $\Delta t$ shall be chosen by cross validation, but in practice, if the query arrival rate is not expected to vary drastically over time, using a time step in minutes yields reasonably accurate arrival models.

It is not efficient to directly solve the optimization problem~\eqref{NHPP formulation w/ periodicity}. Therefore, we design a quadratically approximated ADMM algorithm to efficiently solve it. Introducing auxiliary variables $\mathbf{y}=D^2\mathbf{r}$ and $\mathbf{z}=D_L\mathbf{r}$, we have the augmented Lagrangian for \eqref{NHPP formulation w/ periodicity}
\begin{align*}
&L_{\rho}(\mathbf{r}, \mathbf{y}, \mathbf{z};\bm{\nu}_y,\bm{\nu}_z)=-\mathbf{Q}^T\mathbf{r} + \Delta t\cdot \mathbf{1}^T e^{\mathbf{r}}+\beta_1 \Vert \mathbf{y} \Vert_1 + \frac{\beta_2}{2} \Vert \mathbf{z} \Vert_2^2 \\
&+ \!\bm{\nu}_{y}^T(\mathbf{y}\!-\!D^2\mathbf{r}) \!+\!\bm{\nu}_{z}^T(\mathbf{z}\!-\!D_L\mathbf{r})
\!+\!\frac{\rho}{2}\Vert \mathbf{y} \!-\! D^2\mathbf{r} \Vert_2^2 \!+\! \frac{\rho}{2}\Vert \mathbf{z} \!-\! D_L\mathbf{r} \Vert_2^2
\end{align*}
where $\rho>0$ is the penalty parameter, and $\bm{\nu}_y$, $\bm{\nu}_z$ are dual variables.
Note that solutions to the subproblems with respect to $\mathbf{y}$ and $\mathbf{z}$ can be computed from the proximal operators of $\Vert \cdot \Vert_1$ and $\Vert \cdot \Vert_2^2$ respectively in closed form. The subproblem with respect to $\mathbf{r}$, however, does not admit a closed-form solution due to the exponential term in the loss.
To ensure fast ADMM iterations, we leverage linearized ADMM to solve a quadratically approximated subproblem which can be efficiently solved. Specifically, given the $k$-th iterate $(\mathbf{r}_k, \mathbf{y}_k, \mathbf{z}_k, \bm{\nu}_{y,k}, \bm{\nu}_{z,k})$, we consider the second order Talyor approximation to the exponential term in the augmented Lagrangian
\begin{align*}
&\widehat{L}_{\rho}(\mathbf{r}, \mathbf{y}, \mathbf{z};\bm{\nu}_y,\bm{\nu}_z)
= L_{\rho}(\mathbf{r}, \mathbf{y}, \mathbf{z};\bm{\nu}_y,\bm{\nu}_z) \!-\! \Delta t \!\cdot\! \mathbf{1}^T \!\! e^{\mathbf{r}} \!+\! \Delta t \!\cdot\! \mathbf{1}^T \!\! e^{\mathbf{r}_k}\\
&\hspace{3ex} +\Delta t \cdot (e^{\mathbf{r}_k})^T(\mathbf{r}-\mathbf{r}_k) + \frac{\Delta t}{2}(\mathbf{r}-\mathbf{r}_k)^T\mathrm{diag}(e^{\mathbf{r}_k})(\mathbf{r}-\mathbf{r}_k)
\end{align*}
where $\mathrm{diag}(e^{\mathbf{r}_k})$ denotes the diagonal matrix with $e^{\mathbf{r}_k}$ on the diagonal, and update the iterate $\mathbf{r}_{k+1}=\argmin_{\mathbf{r}}\ \widehat{L}_{\rho}(\mathbf{r}, \mathbf{y}, \mathbf{z};\bm{\nu}_y,\bm{\nu}_z)$ which reduces to solving a sparse linear system.
There are other ways to approximate the subproblem (e.g., by linearizing the proximal penalty only or both the loss and the proximal penalty), but this particular way performs empirically the best for our problem.
The complete ADMM scheme is summarized in Algorithm \ref{algo:admm}, where the ``$\mathrm{SoftThreshold}$'' in line $3$ is the soft thresholding operator defined as $\mathrm{SoftThreshold}(x,c):=\mathrm{sign}(x)\max(\lvert x\rvert -c,0)$.

\begin{algorithm}[t]
\caption{ADMM for solving \eqref{NHPP formulation w/ periodicity}}
\label{algo:admm}
\begin{algorithmic}[1]
\STATEx \textbf{input}: initial guess $\mathbf{r}_0$, $\bm{\nu}_{y,0}$, $\bm{\nu}_{z,0}$, $\mathbf{y}_0=D^2\mathbf{r}_0$, $\mathbf{z}_0=D_L\mathbf{r}_0$
\FOR{$k=0,1,2,\cdots$ until convergence}
\STATE update $\mathbf{r}_{k+1}=A_k^{-1}B_k$, where
\begin{align*}
    &A_k=\Delta t\mathrm{diag}(e^{\mathbf{r}_k})+\rho{D^2}^TD^2 + \rho{D_L}^TD_L\\
    &B_k=\mathbf{Q}-\Delta t e^{\mathbf{r}_k} +\Delta t \mathrm{diag}(e^{\mathbf{r}_k})\mathbf{r}_k + \\
    &\hspace{5ex}{D^2}^T(\bm{\nu}_{y,k} + \rho \mathbf{y}_k) + {D_L}^T(\bm{\nu}_{z,k}+\rho \mathbf{z}_k)
\end{align*}
\STATE update $\mathbf{y}_{k+1}=\mathrm{SoftThreshold}(D^2\mathbf{r}_{k+1}-\frac{1}{\rho}\bm{\nu}_{y,k}, \frac{\beta_1}{\rho})$
\STATE update $\mathbf{z}_{k+1}=\frac{1}{\beta_2+\rho}(\rho D_L \mathbf{r}_{k+1}-\bm{\nu}_{z,k})$
\STATE update $\bm{\nu}_{y,k+1}=\bm{\nu}_{y,k}+\rho(\mathbf{y}_{k+1}-D^2\mathbf{r}_{k+1})$
\STATE update $\bm{\nu}_{z,k+1}=\bm{\nu}_{z,k}+\rho(\mathbf{z}_{k+1}-D_L\mathbf{r}_{k+1})$
\ENDFOR
\STATEx \textbf{output}: $\mathbf{r}_{k+1}$
\end{algorithmic}
\end{algorithm}

We remark on the computational complexity of the ADMM iteration. The bottleneck of each iteration in Algorithm \ref{algo:admm} lies in solving the linear system $A_{k}^{-1}B_k$. The size of the matrix $A_k$ is the same as the length of QPS time series, i.e., $T$, and note that $A_k$ is sparse banded with a bandwidth of order $O(L)$ ($L$ being the length of period) and a total of $O(T)$ non-zero entries due to the structure of the differential matrices $D^2$ and $D_L$. Therefore, the general computational cost in solving $A_{k}^{-1}B_k$ is $O(TL^2)$ (see Section 2.4 in \cite{rue2005gaussian}). In our experiments below, $T$ is in tens of thousands and the linear system can be solved fairly fast.



\section{Scaling Decision}\label{sec:scaling decision}
This section elaborates on the last module of our framework, where the final scaling plan is derived from the predicted traffic dynamics. We first introduce the QoS and cost metrics and their trade-off, and then present the optimization formulations (\eqref{opt:service-constrained}, \eqref{opt:service-constrained using RT}, and 
\eqref{opt:cost-constrained}) and the final scaling scheme Algorithm \ref{sequential scaling:count}.
\subsection{The QoS-Cost Trade-Off}
We measure QoS by two criteria:

\textit{Response Time (RT)}: The RT of each query is the time span between its arrival and the completion of its processing as
\begin{equation*}
\mathrm{RT}_i\!\!=\!\!\begin{cases} 
\!\tau_i + s_i,&\text{if }\xi_i<x_i\text{ (instance not created)}\\
\!x_i\!+\!\tau_i\!+\!s_i\!-\!\xi_i,&\text{if }x_i\!\leq\! \xi_i\!\leq\! x_i\!+\!\tau_i\text{ (instance pending)}\\
\!s_i,&\text{if }x_i+\tau_i< \xi_i\text{ (instance ready)}
\end{cases}
\end{equation*}
or equivalently in a more compact form
$\mathrm{RT}_i=s_i+(\tau_i-(\xi_i-x_i)_+)_+$, 
where $(\cdot)_+:=\max(\cdot, 0)$. $s_i$ is the irreducible processing time and $(\tau_i-(\xi_i-x_i)_+)_+$ is the waiting time  that can be controlled by choosing $x_i$. We shall use the expected RT $E[\mathrm{RT}_i]=\mu_{s} + E[(\tau_i-(\xi_i-x_i)_+)_+]$ as the first criterion.

\textit{Hitting Probability (HP)}: The HP of a query is the probability that the corresponding instance is ready for processing upon arrival. Formally, the HP of $i$-th query is
$\mathrm{HP}_i=P(\xi_i> x_i+\tau_i)$.

From a cloud user's perspective, RT seems more tangible than HP, and here HP is considered as an alternative in that it is a good proxy for RT and as a metric is more universal as the range is always $[0,1]$. In particular, note that both HP and the expected RT are monotone in $x_i$, and when the pending time $\tau_i=\mu_{\tau}$ is a constant, it is not hard to see that $E[\mathrm{RT}_i]\leq \mu_s + \mu_{\tau} (1-\mathrm{HP}_i)$.
Therefore the expected RT can be controlled below a given threshold by making the HP higher than a corresponding level.


We measure the resource cost of an instance by its lifecycle length, i.e., the time span between its creation and deletion. The total resource cost is the total length of all instances' lifecycles. To calculate the lifecycle length, note that if the query arrives before the instance becomes ready then the instance starts processing the query immediately after it is ready, hence the lifecycle length in this case is the pending time plus the processing time, i.e., $\tau_i+s_i$; Otherwise the instance stays idle for a certain amount of time until the query arrives, leading to a lifecycle length $\xi_i+s_i-x_i$. Putting both cases together gives the cost of the $i$-th instance as $\mathrm{cost}_i=(\xi_i-x_i-\tau_i)_+ +\tau_i+s_i$
where $\tau_i+s_i$ is the irreducible fixed cost of pending and processing times, and the first term is the idling time. Note that an earlier instance creation time $x_i$ improves the QoS but increases the cost $\mathrm{cost}_i$. Therefore, the key challenge lies in balancing the fundamental trade-off between QoS improvement and cost reduction.


\subsection{Optimization Formulations}\label{sec:opt formulations}\vspace{-1mm}
We use a constrained optimization to formally study the aforementioned trade-off.
Assuming the current time is zero, we consider the planning of the next $K$ instances.

\subsubsection{QoS-constrained formulation}
Using HP as the QoS criterion, we can specify a desired service level $1-\alpha$, and consider minimizing the expected total cost of the next $K$ instances subject to each query's HP being above $1-\alpha$
\begin{equation}\label{opt:service-constrained}
\begin{aligned}
\min_{x_1,\ldots,x_K\geq 0}&\ \sum_{i=1}^KE[(\xi_i-\tau_i-x_i)_+],~i=1,\ldots,K\\
\mathrm{s.t.}&\ P(\xi_i> x_i+\tau_i)\geq 1-\alpha
\end{aligned}
\end{equation}
where the fixed cost $\mu_{\tau}+\mu_{s}$ is ignored in the objective for simplicity. The stochastic constraint here is a chance constraint which is notiriously hard to solve because of non-convexity \cite{birge2011introduction}. Fortunately, this problem is separable into $K$ individual subproblems, each having a single decision variable $\min_{x_i\geq 0} E[(\xi_i-\tau_i-x_i)_+]\ \mathrm{s.t.}\ P(\xi_i> x_i+\tau_i)\geq 1-\alpha$
which is equivalent to $\max x_i\ \mathrm{s.t.}\ P(\xi_i> x_i+\tau_i)\geq 1-\alpha$ by monotonicity
whose optimal solution is
\begin{equation}\label{sol:service-constrained}
    x_i^* =\alpha \text{ quantile of }(\xi_i-\tau_i).
\end{equation}
Therefore the optimal solution of \eqref{opt:service-constrained} is $(x_1^*,\ldots,x_K^*)$ with each $x_i^*$ given by \eqref{sol:service-constrained}.


Note that, however, the right hand side of \eqref{sol:service-constrained} can be negative, due to the pending time $\tau_i$, in which case a $1-\alpha$ hitting probability is unachievable for the $i$-th query and the problem \eqref{opt:service-constrained} becomes infeasible. This potential infeasibility suggests the necessity of proactively creating the $i$-th instance in advance in order to achieve the desired QoS level.


Alternatively, replacing the HP with the expected RT as the QoS metric yields an optimization with similar structure. In particular, with a user-specified threshold $d$ for the expected RT, the RT-based formulation
\begin{equation}\label{opt:service-constrained using RT}
\begin{aligned}
\min_{x_1,\ldots,x_K\geq 0}&\ \sum_{i=1}^KE[(\xi_i-\tau_i-x_i)_+],~i=1,\ldots,K\\
\mathrm{s.t.}&\ \mu_s + E[(\tau_i-(\xi_i-x_i)_+)_+]\leq d
\end{aligned}
\end{equation}
is also separable and has the optimal solution
\begin{equation}\label{sol:service-constrained using RT}
x_i^*\text{ such that }E[(\tau_i-(\xi_i-x_i^*)_+)_+]= d-\mu_s.
\end{equation}




\subsubsection{Cost-constrained formulation}
If a user prioritizes cost control over QoS, a cost budget $B$ can be specified and we flip the objective and constraint in \eqref{opt:service-constrained using RT} to solve
\begin{equation}\label{opt:cost-constrained}
\begin{aligned}
\min_{x_1,\ldots,x_K\geq 0}&\ \sum_{i=1}^KE[(\tau_i-(\xi_i-x_i)_+)_+],~i=1,\ldots,K\\
\mathrm{s.t.}&\ E[(\xi_i-\tau_i-x_i)_+] + \mu_{\tau}+\mu_s\leq B
\end{aligned}
\end{equation}
where the fixed processing time $\mu_s$ is ignored in the response time objective. Again this problem is separable as for \eqref{opt:service-constrained} and
its optimal solution is
\begin{equation}\label{sol:cost-constrained}
\begin{cases}
x_i^*=0, \text{ \ \ if }E[(\xi_i-\tau_i)_+]\leq B - \mu_{\tau}-\mu_s\\
x_i^*\text{ s.t. }E[(\xi_i-\tau_i-x_i^*)_+]= B- \mu_{\tau}-\mu_s\text{ \ \ otherwise}
\end{cases}.
\end{equation}


\textit{Solution Method and Complexity}: Solving \eqref{sol:service-constrained}, \eqref{sol:service-constrained using RT} and \eqref{sol:cost-constrained} for the optimal scaling decisions boils down to stochastic root finding problems, which in general can be approximated via Monte Carlo sampling. We provide a sort-and-search algorithm (Algorithm \ref{algo:monte carlo}) for approximately solving \eqref{sol:service-constrained using RT}, and similar ones can be designed for \eqref{sol:service-constrained} and \eqref{sol:cost-constrained} as well. Suppose a total of $R$ Monte Carlo samples, $\tau_i^r,\xi_i^r,r=1,\ldots,R$ are drawn to form $\hat{E}(x):=\sum_{r=1}^R(\tau_i^r-(\xi_i^r-x)_+)_+/R$ as an approximation of $E[(\tau_i-(\xi_i-x)_+)_+]$. Note that $\hat{E}(x)$ is piece-wise linear and monotonic in $x$ and its slope changes only at $\{\xi_i^r\}_{r=1}^R$ and $\{\xi_i^r-\tau_i^r\}_{r=1}^R$, thus the idea is to iterate over these linear pieces from the left to the right until the piece that contains the value $d-\mu_s$ is reached.
\begin{algorithm}[t]
\caption{Sort-and-Search for solving \eqref{sol:service-constrained using RT}}
\label{algo:monte carlo}
\begin{algorithmic}[1]
\STATEx \textbf{input}: Monte Carlo samples $\tau_i^r,\xi_i^r,r=1,\ldots,R$, $d$, $\mu_s$
\STATE Sort $\{\xi_i^r\}_{r=1}^R$ and $\{\xi_i^r-\tau_i^r\}_{r=1}^R$ into ascending order $(\xi_i^{(1)},\ldots,\xi_i^{(R)})$ and $((\xi_i-\tau_i)^{(1)},\ldots,(\xi_i-\tau_i)^{(R)})$
\STATE Let $r_1=1,r_2=2,sl=\frac{1}{R},\hat{E}_l=\hat{E}_r=0,x_l=x_r=(\xi_i-\tau_i)^{(1)}$
\WHILE{True}
\IF{$r_1>R$ and $r_2>R$}
\STATE $\hat{x}_i^*=\xi_i^{(R)}$ and break
\ELSIF{$r_2>R$ or $\xi_i^{(r_1)}\leq (\xi_i-\tau_i)^{(r_2)}$}
\STATE $sl'=sl-\frac{1}{R}$, $x_l=x_u$, $x_u=\xi_i^{(r_1)}$, $r_1=r_1+1$
\ELSIF{$r_1>R$ or $\xi_i^{(r_1)}> (\xi_i-\tau_i)^{(r_2)}$}
\STATE $sl'=sl+\frac{1}{R}$, $x_l=x_u$, $x_u=(\xi_i-\tau_i)^{(r_2)}$, $r_2=r_2+1$
\ENDIF
\STATE $\hat{E}_l = \hat{E}_r$ and $\hat{E}_r=\hat{E}_r+sl(x_u-x_l)$
\IF{$\hat{E}_l\leq d-\mu_s \leq \hat{E}_r$}
\STATE $\hat{x}_i^*=(d-\mu_s-\hat{E}_l)/sl+x_l$ and break
\ENDIF
\STATE $sl=sl'$
\ENDWHILE
\STATEx \textbf{output}: $\hat{x}_i^*$
\end{algorithmic}
\end{algorithm}
Specifically, steps 4-10 of Algorithm \ref{algo:monte carlo} update the slope at each break point (minus $\frac{1}{R}$ when passing a $\xi_i^r$, plus $\frac{1}{R}$ when passing a $\xi_i^r-\tau_i^r$), step 11 updates $\hat{E}(x)$ at the endpoints $x_l,x_u$ using the slope in $O(1)$ time rather than the direct summation that costs $O(R)$ time. Since the sorting step can be done in $O(R\log R)$ time, and the search step has a linear time complexity, therefore the overall complexity of Algorithm \ref{algo:monte carlo} is $O(R\log R)$. The number of actions $K$ to consider usually scales linearly with the QPS of the traffic, hence the practical computational cost is roughly $O(\text{QPS}\cdot R\log R)$. The linear complexity in both QPS and sampling size makes our method highly scalable both in theory and practice.

\vspace{-1mm}
\subsection{Proactive Strategies and QoS Guarantees}\label{sec:prob guarantees}\vspace{-1mm}
As briefly mentioned in the last subsection, proactively scaling out instances is necessary for consistently maintaining a desired level of QoS, and this subsection presents a novel proactive strategy. We will use HP as our QoS criterion.

We explain the main idea of our proactive strategies. A naive strategy based on the solution \eqref{sol:service-constrained} is to plan the instance creation time at the current time for a batch of $K$ upcoming queries, and then wait until all planned $K$ instances are consumed to do planning for the next batch of $K$ queries. A key issue of this naive strategy however is that we may not be able to have the instances ready in time for the first few queries in each batch, because all previously created instances have been consumed at the time of planning. Therefore, a crucial step is to start planning for the next batch of queries once the number of instances left reaches a certain carefully chosen threshold $\kappa$, rather than after all are used, so that the first few upcoming queries are taken care of by the left instances and only queries that arrive further later need to be included in planning. Algorithm \ref{sequential scaling:count} details our proposal.
\begin{algorithm}[t]
\caption{Sequential scaling: $\mathrm{HP}$-constrained}
\label{sequential scaling:count}
\begin{algorithmic}[1]
\STATEx \textbf{input:} intensity $\lambda_t$ for $t\geq 0$, an upper bound $\overline{\lambda}$ for $\lambda_t\leq \overline{\lambda}$, target hitting probability $1-\alpha$, planning frequency $m\geq 1$
\STATEx \textbf{initialization:} total number of queries seen so far $N=0$
\STATE Compute the threshold
\begin{equation}\label{remaining instance threshold}
    \kappa:=\max\{i:\alpha \text{ quantile of }({\gamma_i}/{\overline{\lambda}}-\tau_i)<0, i\geq 1\}
\end{equation}
where $\gamma_i,\tau_i$ are independent, and $\gamma_i$ follows the Gamma distribution with shape $i$ and scale $1$
\REPEAT
\STATEx \textit{Planning Step:}
\IF{$N=0$}
\STATE Compute instance creation time $\![x_1^*,\!\ldots\!,x_{\kappa+m}^*]\!$ by \eqref{sol:service-constrained}
\ELSE
\STATE Given the arrival history so far, compute instance creation time $\![x_{N+\kappa+1}^*,\!\ldots\!,x_{N+\kappa+m}^*]\!$ by \eqref{sol:service-constrained}
\ENDIF
\STATEx \textit{Execution Step:}
\STATE Execute Alg. \ref{algo:dynamics} until the query's arrival time $\xi_{N+m}$ 
\STATE $N= N+m$
\UNTIL{forever}
\end{algorithmic}
\end{algorithm}

To explain Algorithm \ref{sequential scaling:count}, in line $1$ the aforementioned threshold $\kappa$ is calculated in \eqref{remaining instance threshold} by assuming the queries arrive according to a constant intensity $\overline{\lambda}$ that upper bounds the true $\lambda_t$. According to the way $\kappa$ is computed and the discussion in Section \ref{sec:opt formulations}, if the query arrivals follow the constant intensity $\overline{\lambda}$, for each $k>\kappa$, one is able to get the instance ready in time for the $k$-th upcoming query to achieve a $1-\alpha$ hitting probability. Since the actual $\lambda_t
\leq \overline{\lambda}$, the $k$-th upcoming query under the actual intensity always arrives later than that under $\overline{\lambda}$, and thus the $1-\alpha$ hitting probability must be attainable under the actual intensity too. Lines $2$-$10$ then updates the instance creation plan every $m$ query arrivals. Note that in each round of planning we only need to calculate instance creation times for the $(\kappa+1)$-th to $(\kappa+m)$-th upcoming queries because the first $\kappa$ upcoming queries already have their instances scheduled in the last round. This way, we always plan instance creation times at least $\kappa+1$ arrivals ahead, therefore the desired $1-\alpha$ hitting probability can be achieved for every single arrival.


We provide two propositions below on the hitting probability achieved by Algorithm \ref{sequential scaling:count} under different conditions. Proposition \ref{prop:HP guarantee} shows that the hitting probability is exactly $1-\alpha$ when the underlying arrival process is an NHPP and the intensity is known, serving as an ideal case for Algorithm \ref{sequential scaling:count}. In practice, however, the intensity needs to be predicted or estimated with a certain error, and in this case Proposition \ref{prop:HP error} shows that the hitting probability error grows at most linearly in the relative error of the intensity estimate that is used to make scaling decisions, therefore the attained hitting probability quickly approaches $1-\alpha$ as the intensity estimate becomes more accurate. They are formally stated as:
\begin{proposition}\label{prop:HP guarantee}
If the query arrivals $\{\xi_i:i\geq 1\}$ follow an NHPP with intensity $\lambda_t,t\geq 0$ and Algorithm \ref{sequential scaling:count} is used with $\lambda_t$ as an input, then $P(\xi_i\geq x_i^*+\tau_i)=1-\alpha$ for every $i> \kappa$. Moreover, for every $N > \kappa$ the variance of the hitting ratio $\mathrm{Var}(\frac{1}{N-\kappa}\sum_{i=\kappa+1}^N\mathbf{1}(\xi_i\geq x_i^*+\tau_i))\leq \frac{2(\kappa+m)\alpha(1-\alpha)}{N-\kappa}$.
\end{proposition}
Note that Proposition \ref{prop:HP guarantee} also shows that the empirical hitting ratio has a quickly diminishing variance as the queries accumulate and hence should be close to the target $1-\alpha$.
\begin{proposition}\label{prop:HP error}
Suppose that the queries arrive according to an NHPP with unknown intensity $\lambda^*_t$, and that Algorithm \ref{sequential scaling:count} is used with an estimate $\lambda_t$ such that $\lvert \lambda_t - \lambda_t^* \rvert\leq \epsilon\lambda_t^*$ for some small $\epsilon>0$, then for each $i>\kappa$ the hitting probability
\begin{equation*}
    \lvert P(\xi_i\geq x_i^*+\tau_i) - (1-\alpha) \rvert\leq \frac{\epsilon}{1-\epsilon}(q_{\kappa+m,\alpha} + \mu_{\tau}\sup_{t}\lambda_t)\vspace{-2mm}
\end{equation*}
where $q_{\kappa+m,\alpha}$ is the $\alpha$-quantile of the Gamma distribution with shape $\kappa+m$ and scale $1$.
\end{proposition}
We sketch the main ideas in proving the two results. In Proposition \ref{prop:HP guarantee}, the exact $1-\alpha$ hitting probability can be established because scaling for each incoming query is planned sufficiently ahead of time as explained before. The variance of the hitting ratio of the first $N$ queries can be controlled because scaling is planned at most $\kappa+m$ steps ahead and hence by the independent-increment property of a Poisson process the hitting events of two queries are statistically independent whenever they are at least $\kappa+m$ apart. This is why the usual reciprocal relationship of variance and sample size $N$ shows up with the additional factor $\kappa+m$ in the numerator. Proposition 2 is proved by first representing the hitting probability in terms of Gamma distribution functions through a time rescaling of the NHPP into a homogeneous one, so that the error analysis of hitting probabilities boils down to that of the associated integrated intensities, and then bounding the error of the latter.

The detailed proofs are as follows:
\vspace{-3mm}
\begin{proof}[Proof for Proposition \ref{prop:HP guarantee}]
Denote by $\mathcal H_{t}$ the arrival history, i.e., a list of times of arrivals, up to time $t$, and by $h_i:=\mathbf{1}(\xi_i\geq x_i^*+\tau_i)$. For each $i>\kappa$, let $c_i$ be the integer such that $\kappa+c_im<i\leq \kappa+(c_i+1)m$, then the design of Algorithm \ref{sequential scaling:count} ensures that $x_i^*$ is planned when the $c_im$-th query arrives, therefore by the way $x_i^*$ is calculated in \eqref{sol:cost-constrained}, $P(\xi_i\geq x_i^*+\tau_i) = E[P(\xi_i\geq x_i^*+\tau_i\vert \mathcal{H}_{\xi_{c_im}})] = E[1-\alpha]=1-\alpha$.

To calculate the variance of the empirical hitting ratio, we first note that, because of the independent-increment property of NHPP and that the horizon of the scaling plan is $\kappa+m$, $h_i$ and $h_j$ are independent whenever $\lvert i-j
\rvert\geq \kappa+m$. Secondly, by Cauchy-Schwartz inequality, $\mathrm{Cov}(h_i, h_j)\leq \sqrt{\mathrm{Var}(h_i)\mathrm{Var}(h_j)}=\alpha(1-\alpha)$. Therefore, we can write
\begin{align*}
&\mathrm{Var}\Big(\!\sum_{i=\kappa+1}^N\!\!\mathbf{1}(\xi_i\geq x_i^*+\tau_i)\Big)
= \!\! \sum_{i=\kappa+1}^N\!\!\mathrm{Var}(h_i)\!+\!\sum_{i\neq j}\mathrm{Cov}(h_i, h_j)\\
&\leq(N - \kappa)\alpha(1-\alpha) + \sum_{i=\kappa+1}^N\sum_{j:0<\lvert i-j
\rvert< \kappa+m}\mathrm{Cov}(h_i, h_j)\\
&\leq 2(\kappa+m)(N - \kappa)\alpha(1-\alpha),
\end{align*}
and dividing the right hand side by $(N-\kappa)^2$ gives the desired variance bound.
\end{proof}

\vspace{-2mm}
\begin{proof}[Proof for Proposition \ref{prop:HP error}]
It follows from Proposition \ref{prop:HP guarantee} that $P_{\lambda_t}(\xi_i\geq x_i^*+\tau_i) = 1-\alpha$, where the subscript $\lambda_t$ is used to represent the probability when the underlying intensity is $\lambda_t$. Therefore, it suffices to bound the difference $P_{\lambda_t^*}(\xi_i\geq x_i^*+\tau_i) - P_{\lambda_t}(\xi_i\geq x_i^*+\tau_i)$. To proceed, we borrow the symbol $c_i$ from the proof of Proposition \ref{prop:HP guarantee}, and we know that $x_i^*$ is calculated at time $\xi_{c_im}$.
Denote by $\eta = \int_{\xi_{c_im}}^{x_i^*+\tau_i}\lambda_tdt$, $\eta^*=\int_{\xi_{c_im}}^{x_i^*+\tau_i}\lambda_t^*dt$, and $F_k$ the cdf of Gamma distribution with shape parameter $k$ and scale parameter $1$. By conditioning on the instance pending time $\tau_i$ and the history $\mathcal H_{c_im}$, and time-rescaling, we can express the hitting probability as
\begin{eqnarray*}
    &&P_{\lambda_t}(\xi_i\geq x_i^*+\tau_i\vert \tau_i, \mathcal H_{c_im})
    =P\big(\Gamma(i-c_im, 1) \\ &\geq& \int_{\xi_{c_im}}^{x_i^*+\tau_i}\lambda_tdt\vert \tau_i, \mathcal H_{c_im}\big)
    = 1-F_{i-c_im}(\eta)
\end{eqnarray*}
where $\Gamma$ denotes a Gamma random variable, and similarly $P_{\lambda_t^*}(\xi_i\geq x_i^*+\tau_i\vert \tau_i, \mathcal H_{c_im})=1-F_{i-c_im}(\eta^*)$. Note that by the construction of $x_i^*$ we have $E[1-F_{i-c_im}(\eta)\vert \mathcal H_{c_im}]=1-\alpha$. To calculate $E[1-F_{i-c_im}(\eta^*)\vert \mathcal H_{c_im}]$, we need to use the intensity error bound
\begin{equation*}
\lvert \eta^* - \eta \rvert\leq \int_{\xi_{c_im}}^{x_i^*+\tau_i} \lvert \lambda_t^* - \lambda_t \rvert dt\leq \int_{\xi_{c_im}}^{x_i^*+\tau_i} \epsilon \lambda_t^* dt= \epsilon\eta^*\vspace{-2mm}
\end{equation*}
and a simple bound of the Gamma pdf $\frac{1}{(k-1)!}x^{k-1}e^{-x}\leq 1$ for all $x>0$ and all integer $k\geq 1$. So we have
\begin{eqnarray*}
&&\lvert E[1-F_{i-c_im}(\eta^*)\vert \mathcal H_{c_im}] - E[1-F_{i-c_im}(\eta)\vert \mathcal H_{c_im}]\rvert\\
&\leq &E[\lvert F_{i-c_im}(\eta^*)-F_{i-c_im}(\eta)\rvert \vert \mathcal H_{c_im}]\\
&\leq &E[\lvert  \eta - \eta^*\rvert \vert \mathcal H_{c_im}]\leq \epsilon E[\eta^*\vert \mathcal H_{c_im}]\leq \frac{\epsilon}{1-\epsilon} E[\eta\vert \mathcal H_{c_im}]\vspace{-2mm}
\end{eqnarray*}
where the last inequality follows from $\eta^* - \eta \leq \epsilon\eta^*\Rightarrow (1-\epsilon)\eta^*\leq \eta $. To further bound $E[\eta\vert \mathcal H_{c_im}]$, note that $\int_{\xi_{c_im}}^{x_i^*}\lambda_tdt\leq \eta\leq \int_{\xi_{c_im}}^{x_i^*}\lambda_tdt + \tau_i\sup_{t}\lambda_t$. Since $F_{i-c_im}$ is monotone, it holds that $F_{i-c_im}(\int_{\xi_{c_im}}^{x_i^*}\lambda_tdt)\leq E[F_{i-c_im}(\eta)\vert \mathcal H_{c_im}]=\alpha$, implying that $\int_{\xi_{c_im}}^{x_i^*}\lambda_tdt\leq F_{i-c_im}^{-1}(\alpha)\leq F_{\kappa+m}^{-1}(\alpha)$ since $F_{i-c_im}\geq F_{\kappa+m}$, hence $E[\eta\vert \mathcal H_{c_im}]\leq F_{\kappa+m}^{-1}(\alpha) + \mu_{\tau}\sup_{t}\lambda_t$. 
Through Jensen's inequality, we obtain the probability error  
\begin{eqnarray*}
&&\lvert P_{\lambda_t^*}(\xi_i\geq x_i^*+\tau_i) - P_{\lambda_t}(\xi_i\geq x_i^*+\tau_i)\rvert\\
&\leq &E[\lvert P_{\lambda_t^*}(\xi_i\geq x_i^*+\tau_i\vert \mathcal H_{c_im}) \!-\! P_{\lambda_t}(\xi_i\geq x_i^*\!+\!\tau_i\vert \mathcal H_{c_im})\rvert]\\
&\leq &\frac{\epsilon}{1-\epsilon}\big(F_{\kappa+m}^{-1}(\alpha) + \mu_{\tau}\sup_{t}\lambda_t\big).
\end{eqnarray*}
This concludes the error bound in the proposition.
\end{proof}

Some practical guidelines for using Algorithm \ref{sequential scaling:count} are as follows. First, Proposition \ref{prop:HP guarantee} suggests a QoS variance that scales with $\kappa$, thus we suggest using estimates of the local intensity at the current time instead of a global upper bound $\overline{\lambda}$ in \eqref{remaining instance threshold} to obtain a smaller value for $\kappa$ so that the QoS is stabler. Second, the hitting probability can be calibrated for better accuracy by choosing a set of nominal levels $0<p_1<\cdots <p_B<1$ to run Algorithm \ref{sequential scaling:count} on training data so that the resulting actual hitting probabilities $0<\hat p_1<\cdots <\hat p_B<1$ cover a wide range of $[0,1]$. This gives a mapping between the nominal and actual hitting probabilities, which can then be used to pick the right nominal level to ensure a desired actual hitting probability.




\section{Experimental Results}\label{sec:experiments}
We evaluate and compare various proactive autoscaling methods on three real-world datasets to demonstrate the advantages of the proposed RobustScaler. We also evaluate and analyze various aspects, such as scalability, robustness, accuracy, and performance in real environments, of our framework.


\vspace{-1mm}
\subsection{Algorithms, Datasets, and Metrics}\vspace{-1mm}
\subsubsection{Autoscaling Algorithms}\label{sec:autoscaling methods intro}
We consider two heuristic autoscaling strategies and three variants of our RobustScaler:
\begin{itemize}
    \item Backup Pool (BP): This strategy constantly maintains a pool of $B$ instances. Upon each query arrival, one of the instances is used to process the query, and then the pool is immediately replenished with a new instance. Using $B=0$ is equivalent to a purely reactive strategy.
    \item Adaptive Backup Pool (AdapBP): This heuristic is an adaptive version of Backup Pool where the pool size is regularly adjusted according to the QPS level. Specifically, the average QPS during the most recent ten minutes is used as an estimate of the current arrival rate, and every ten minutes the pool size is reset to be the estimate multiplied by a pre-fixed constant.
    
    \item RobustScaler-HP: This is our framework with scaling scheme Algorithm \ref{sequential scaling:count} but with two changes: 1) $\kappa$ is chosen time-dependent according to the local intensity at each planning step to improve cost efficiency, and 2) planning frequency is specified by a fixed time interval of $\Delta $ seconds instead of number of queries, meaning that planning is performed every $\Delta$ seconds and in each round instance creation times that lie within the next $\Delta$ interval are all computed.
    
    \item RobustScaler-RT and RobustScaler-cost: The same as RobustScaler-HP except that the scaling decisions are computed from the solution \eqref{sol:service-constrained using RT} and \eqref{sol:cost-constrained}, respectively.
    
\end{itemize}

Note that many existing autoscaling algorithms rely on system metrics such as CPU and memory utilization, whereas in the scaling-per-query scenario the time of query arrival and its uncertainty are the main factors considered in making scaling decisions, therefore these method cannot handle the scaling-per-query scenario directly, and we do not include them (such as Horizontal Pod Autoscaler (HPA) scheme in Kubernetes~\cite{burns2019kubernetes}) in comparison.

\subsubsection{Datasets}

\begin{figure}[!t]
    \centering
    \includegraphics[width=.9\linewidth]{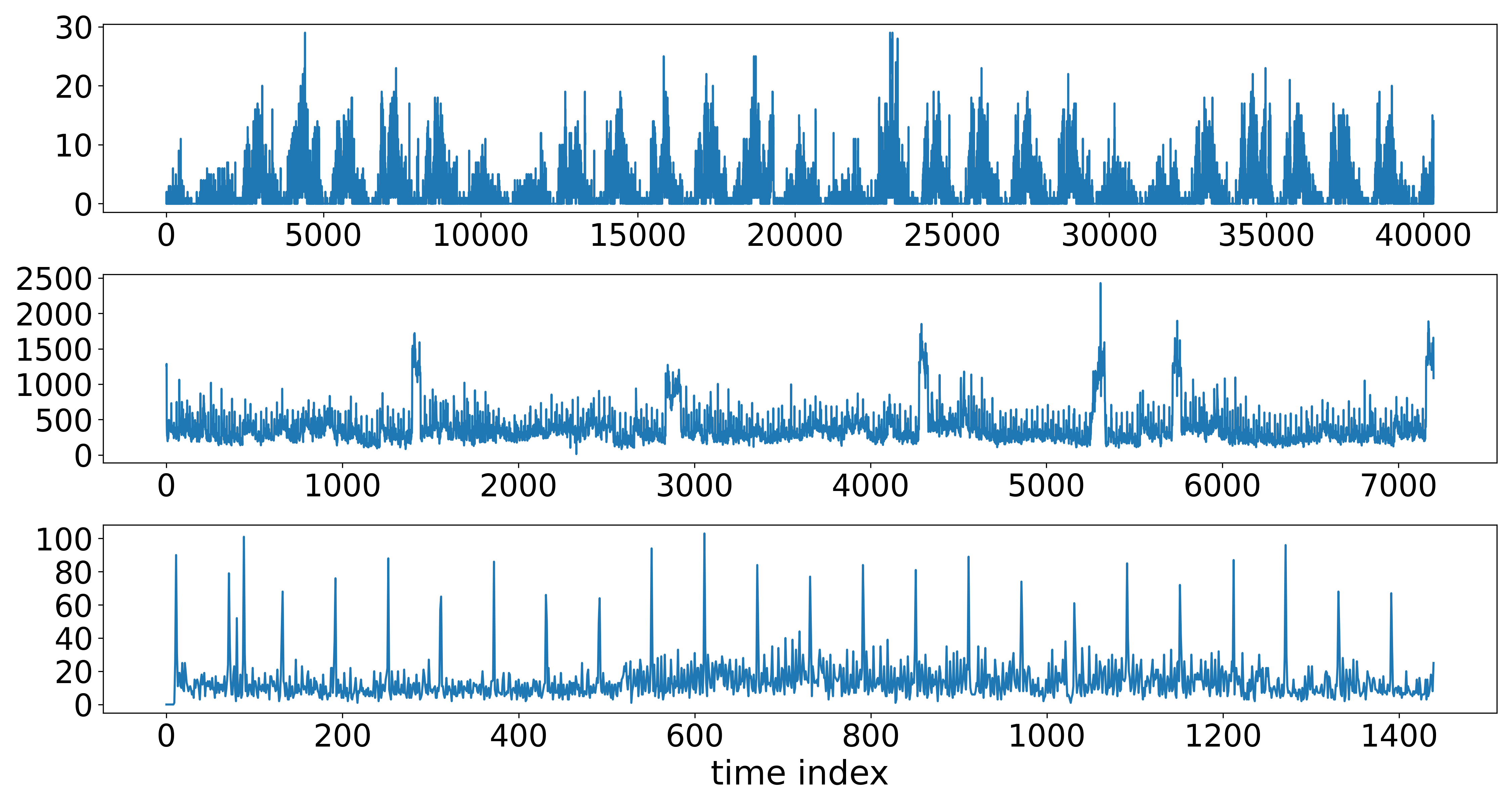}
    \vspace{-4mm}
    \caption{Real-world QPS series, from top to bottom: CRS trace, Alibaba trace, Google trace.}
    \label{fig:real series}
\end{figure}

We test the autoscaling methods on 3 real-world traces. The first trace, ``CRS'', is from the container registry service of a top cloud service provider that contains time information of a total of $21059$ queries for building container images, including start and end times of query processing, over $4$ weeks. The first three weeks of the CRS trace are used as training data and the last week for testing. The second dataset is the Google cluster data 2019 trace~\cite{tirmazi2020borg} that describes workloads running on eight Borg clusters for the month of May 2019. We use the job trace data of ``cluster b'' and use a subset of $24$ hours for the experiments which has $20254$ job queries in total, among which the first $18$ hours of data are training and the last $6$ hours are testing data. The third dataset is the Alibaba Cluster Trace 2018~\cite{guo2019limits} that contains job information of around $4000$ machines over $8$ days. We select $5$ days of data with a total of $503850$ job records, among which the first four days are for training and the last day for testing. We find that the selected parts of Google and Alibaba trace are representative of the original traces in terms of workload patterns and at the same time allow faster experimental runs and discoveries. We replay these real-world traces with various autoscalers to evaluate their performance.
Figure \ref{fig:real series} plots the QPS series of the three traces after aggregating job/query counts with a time resolution $\Delta t=60$ seconds. We can see that the CRS trace is quite noisy because of the relatively low traffic but seems to have a weekly pattern, and that Alibaba trace and Google trace both have recurrent spikes but the Alibaba trace has an unexpected burst/spike on the fourth day which brings challenges to modeling and prediction. Therefore, these datasets are challenging enough for differentiating different autoscaling algorithms. 

\subsubsection{Metrics}
We use the following evaluation metrics:
\begin{itemize}
    \item hit\_rate: the proportion of queries for which at least one instance is ready upon arrival
    \item total\_cost: the sum of lifecycle lengths in seconds of all instances
    \item relative\_cost: ratio of total\_cost versus the cost of the pure reactive BP with $B=0$
    \item rt\_avg: average response time in seconds of all queries.
\end{itemize}

\begin{figure}[!t]
    \centering
    \subfigure[hit\_rate vs relative\_cost on CRS trace]{\includegraphics[width=0.24\textwidth]{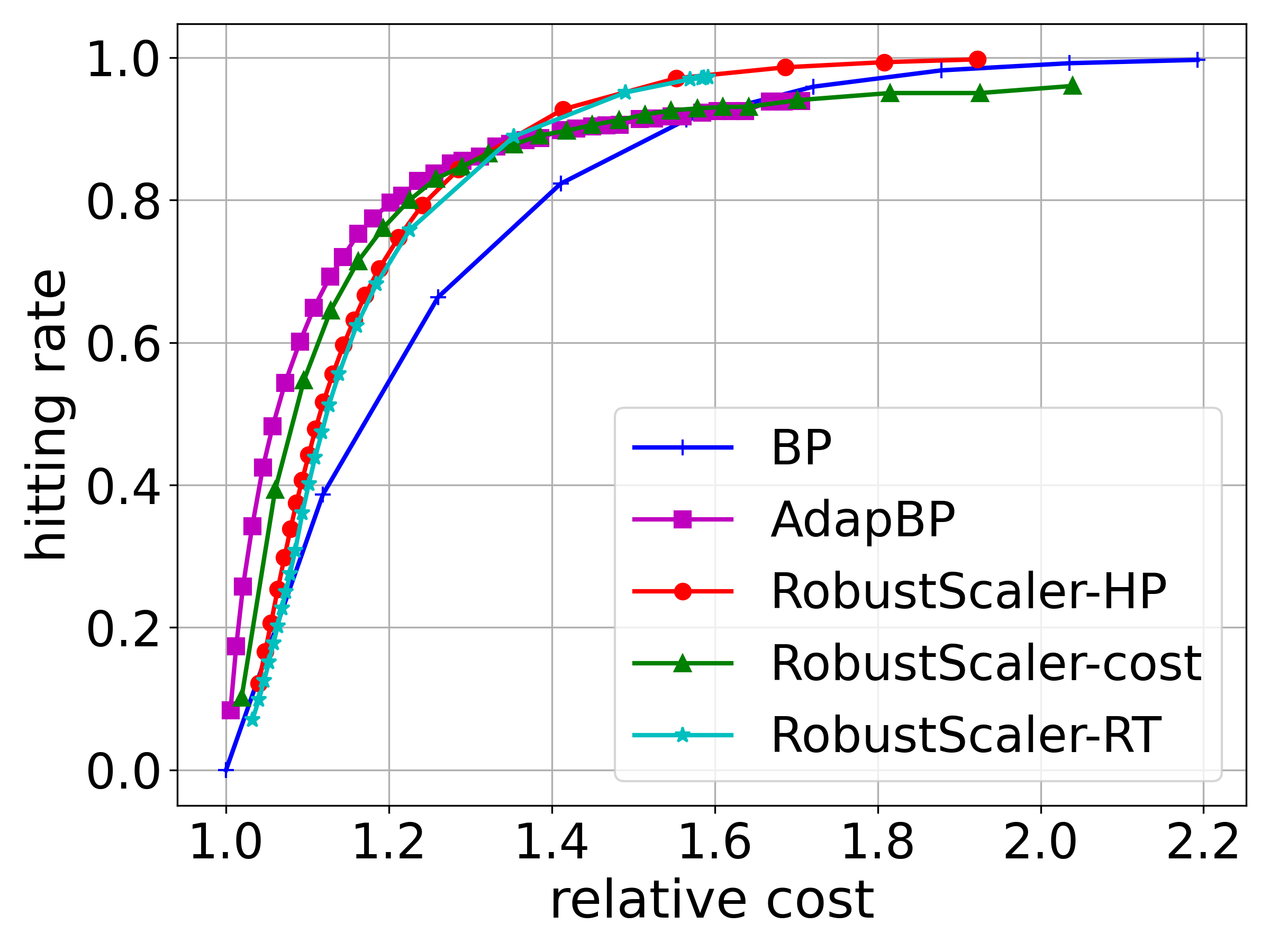}\label{fig:pareto of hit and cost on acr}}
    \subfigure[rt\_avg vs relative\_cost on CRS trace]{\includegraphics[width=0.24\textwidth]{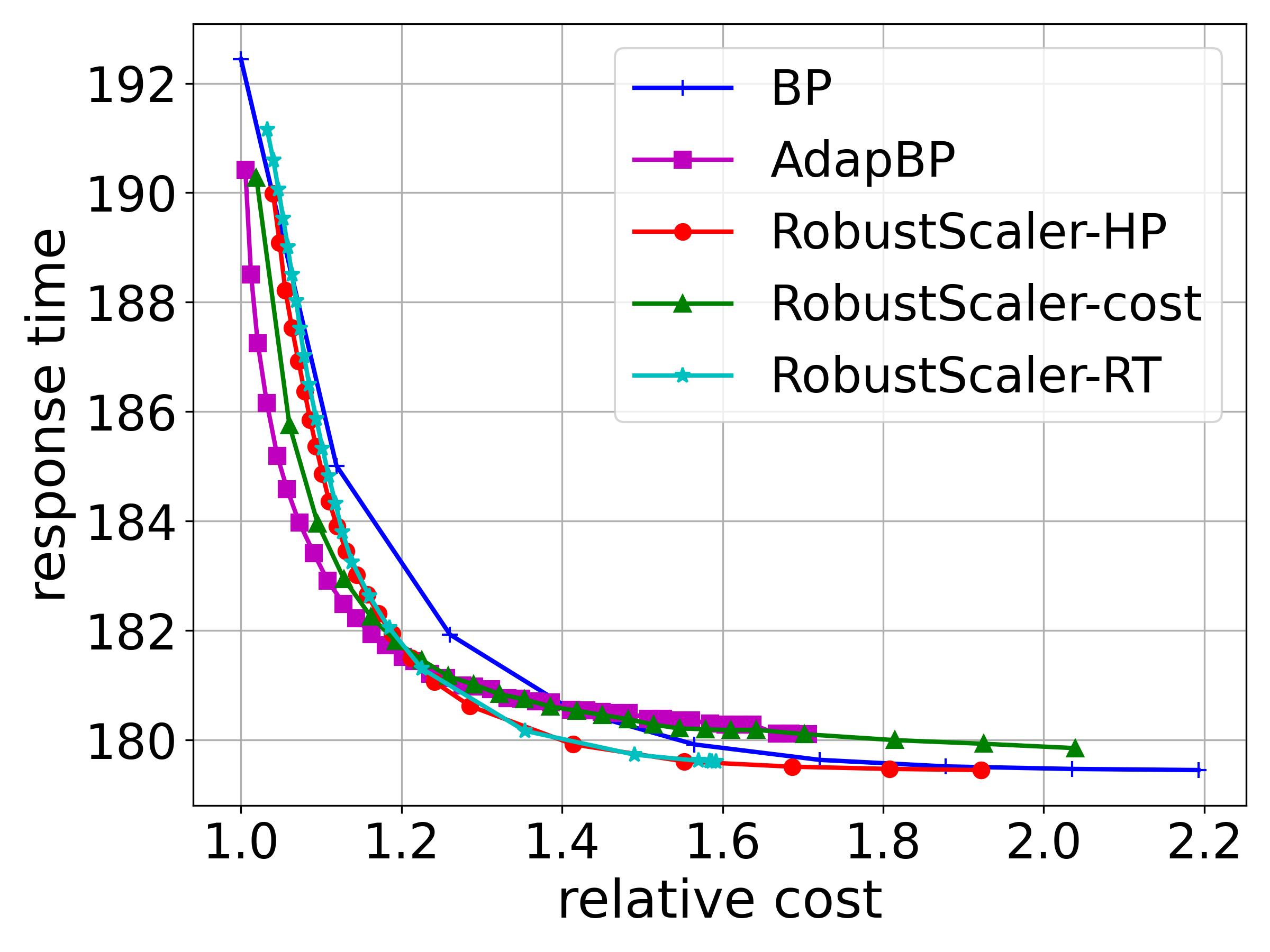}\label{fig:pareto of rt and cost on acr}}\\\vspace{-0.1cm}
    \subfigure[hit\_rate vs relative\_cost on Alibaba trace]{\includegraphics[width=0.24\textwidth]{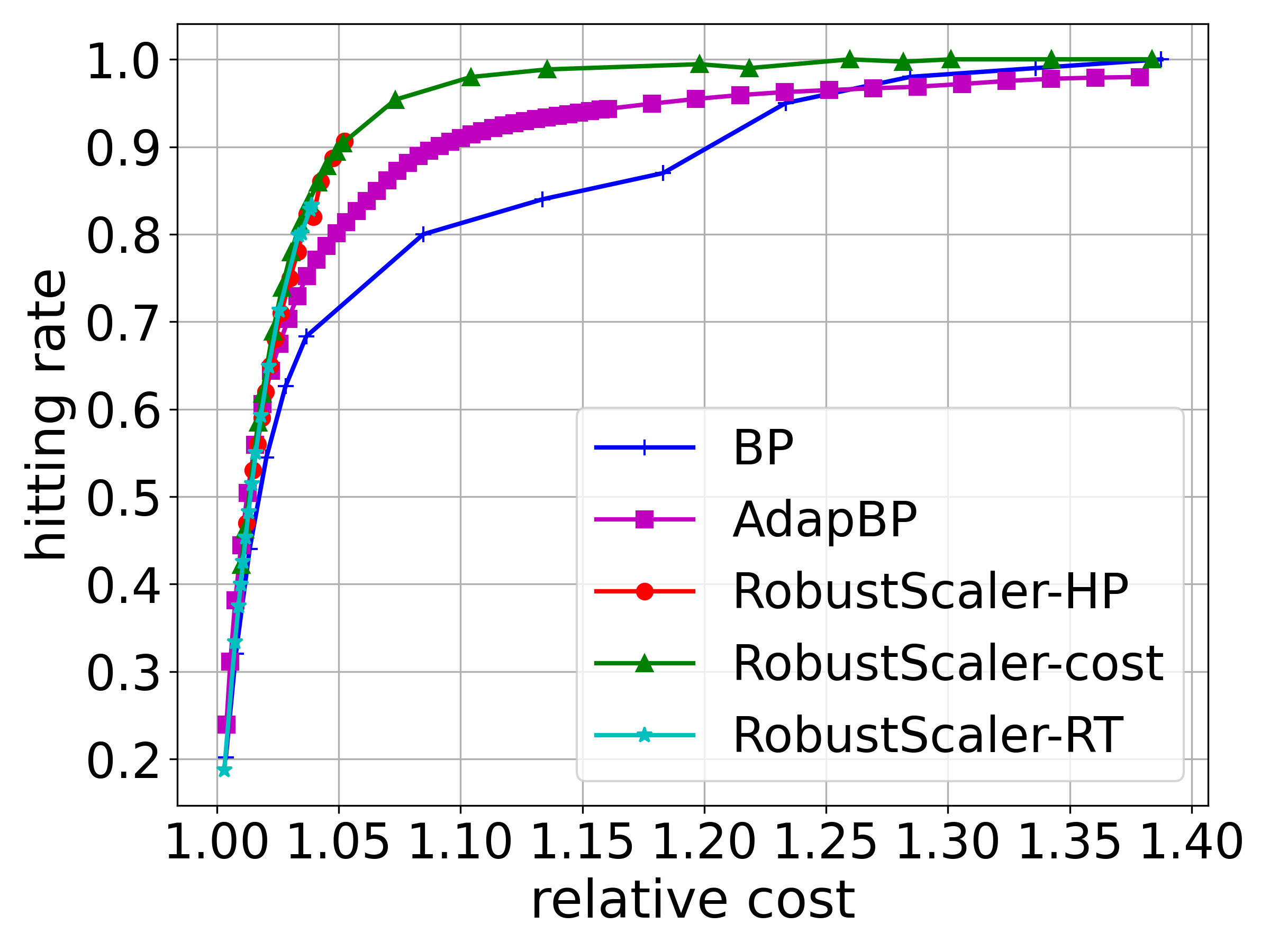}\label{fig:pareto of hit and cost on alibaba}} 
    \subfigure[rt\_avg vs relative\_cost on  Alibaba trace]{\includegraphics[width=0.24\textwidth]{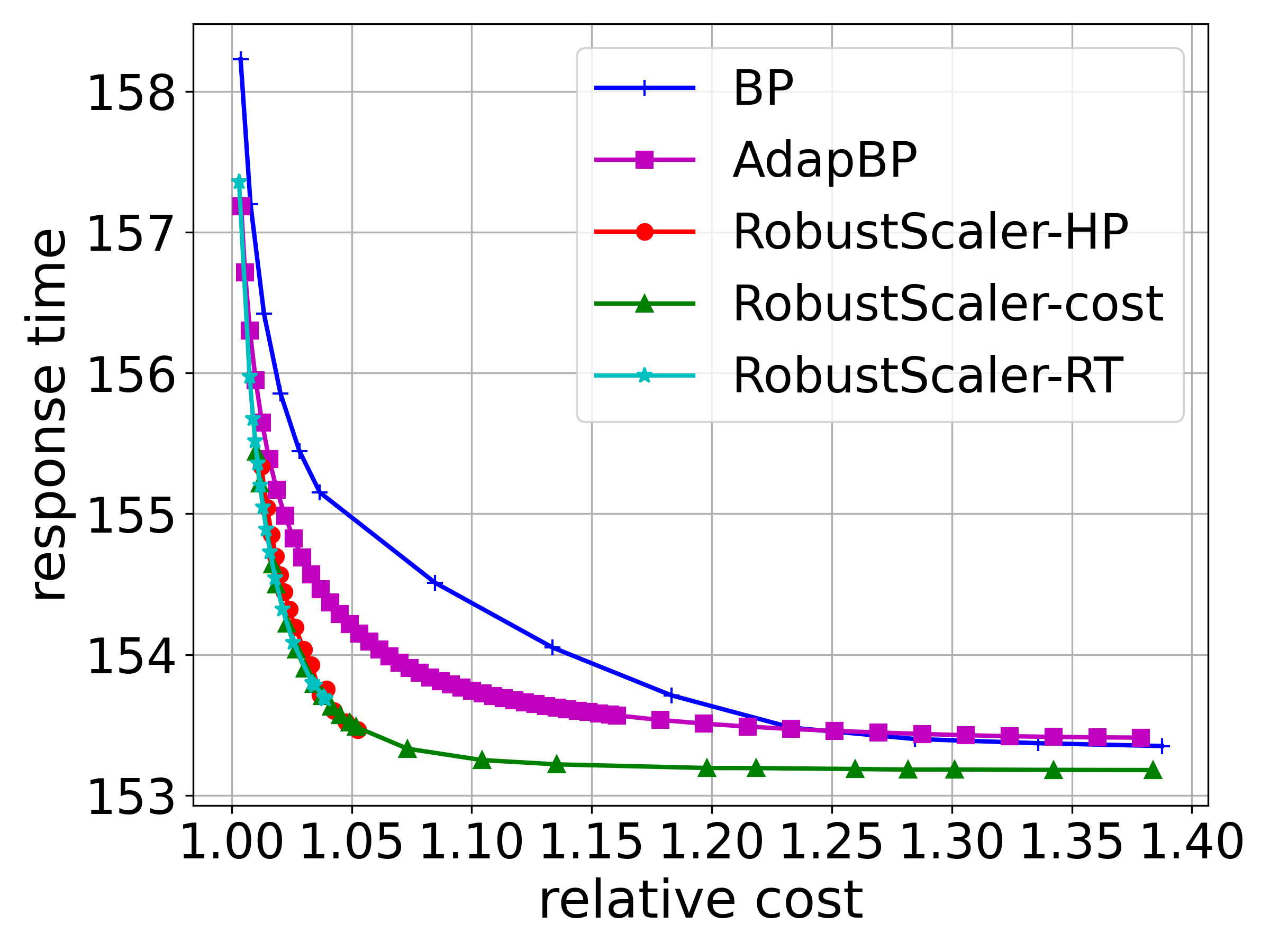}\label{fig:pareto of rt and cost on alibaba}}\\\vspace{-0.1cm}
    \subfigure[hit\_rate vs relative\_cost on Google trace]{\includegraphics[width=0.24\textwidth]{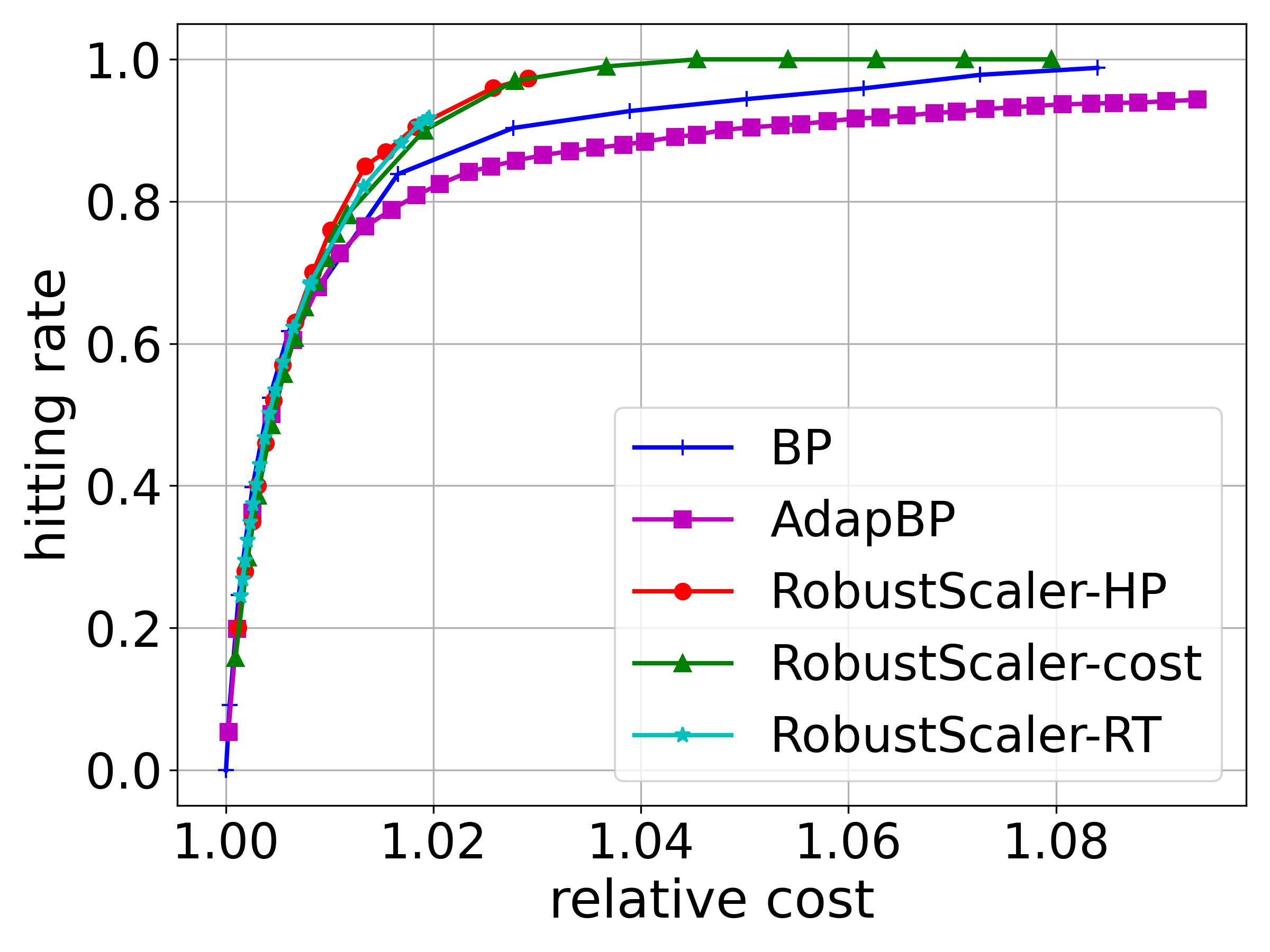}\label{fig:pareto of hit and cost on google}}
    \subfigure[rt\_avg vs relative\_cost on  Google trace]{\includegraphics[width=0.24\textwidth]{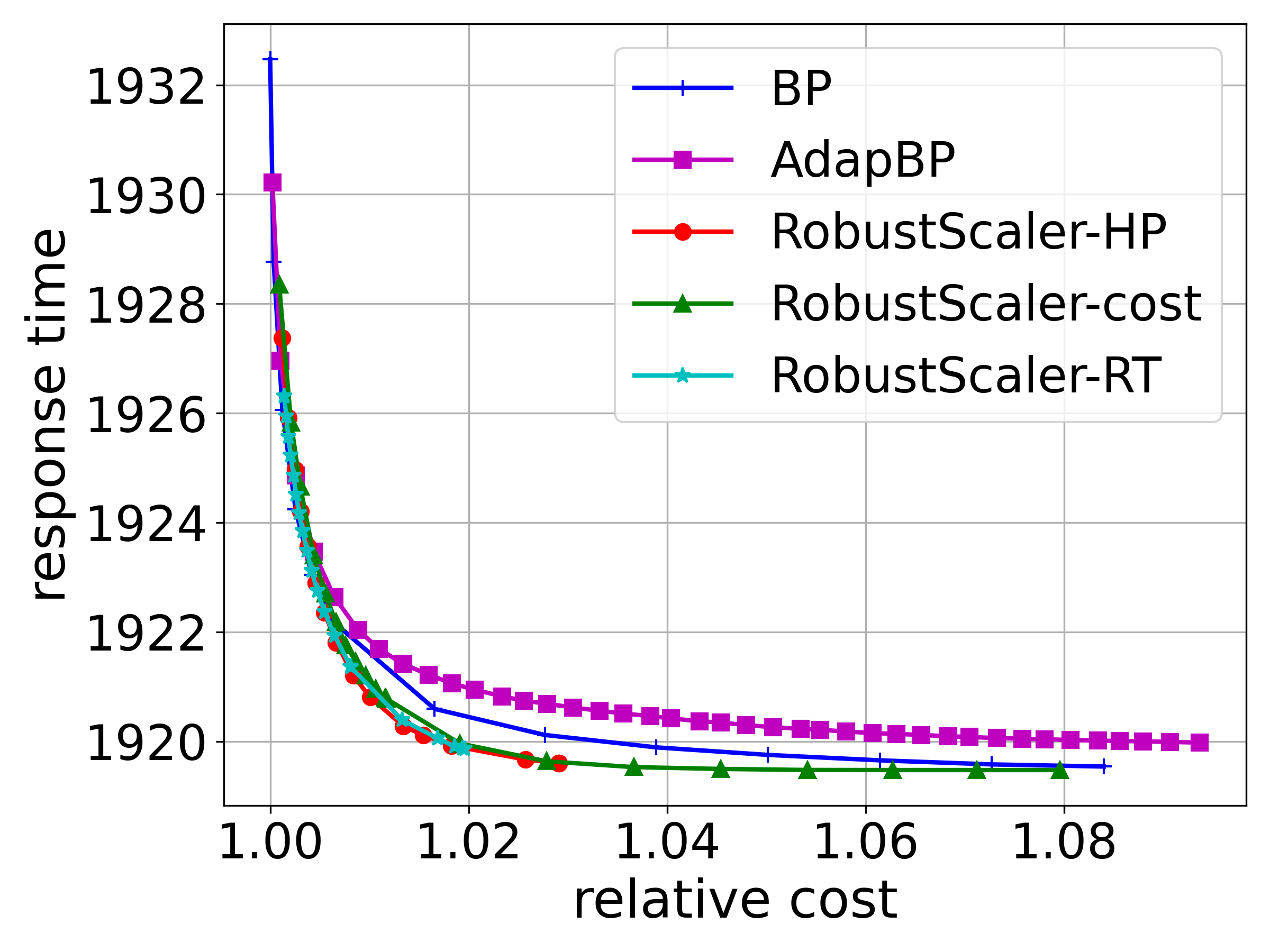}\label{fig:pareto of rt and cost on google}}
    \vspace{-4mm}
    \caption{Pareto plots of different autoscalers.}
    \vspace{-2mm}
    \label{fig:Pareto_plot}
\end{figure}

\begin{figure}[!t]
    \centering
    \subfigure[variance vs mean of hit\_rate]{\includegraphics[width=0.24\textwidth]{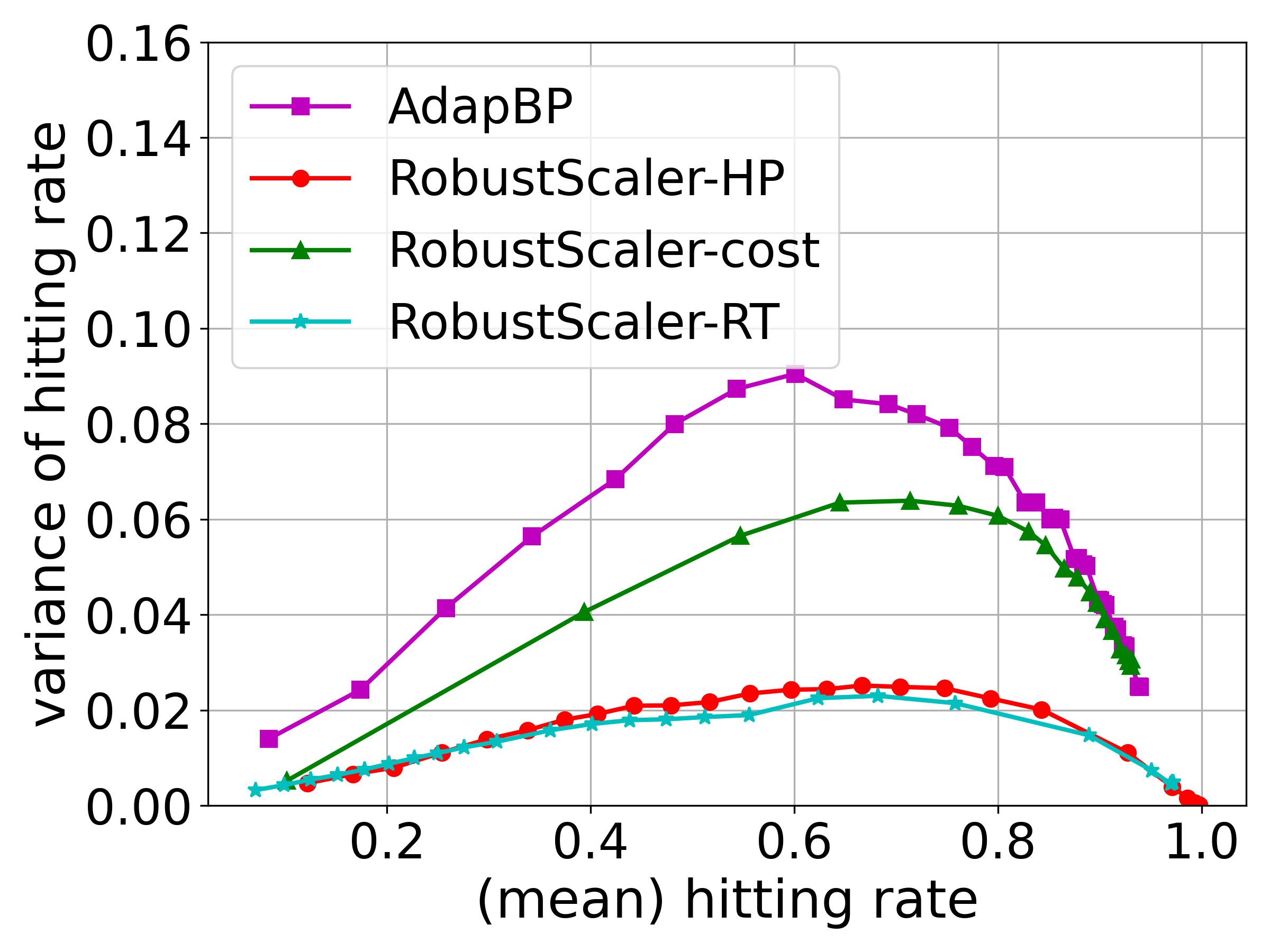}\label{fig:var of hit on acr}}
    \subfigure[variance vs mean of rt\_avg]{\includegraphics[width=0.24\textwidth]{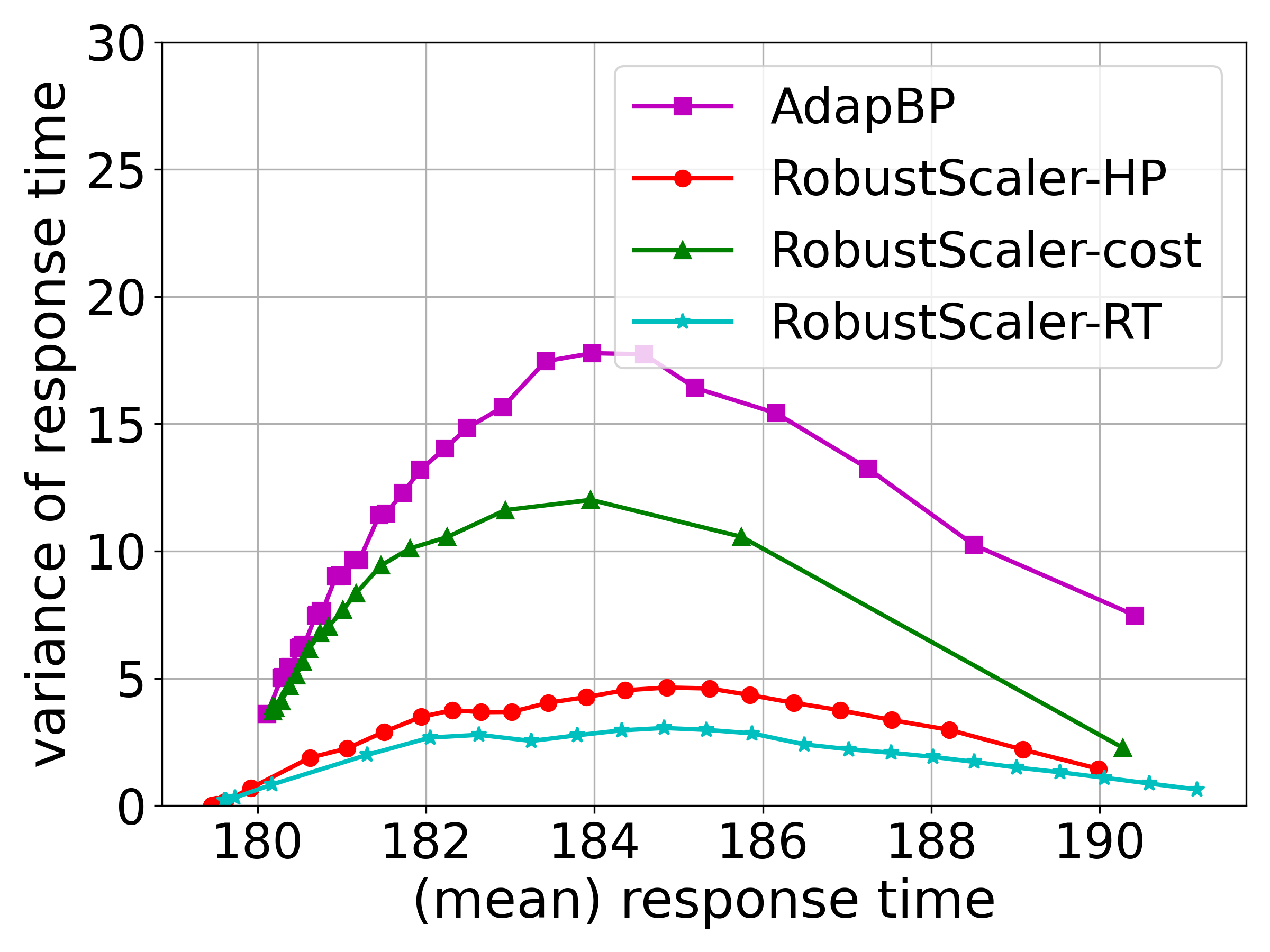}\label{fig:var of rt on acr}}
    \vspace{-4mm}
    \caption{Variance of QoS on CRS trace.}
    \label{fig:var_plot}
\end{figure}

\begin{figure}[!t]
    \centering
    \subfigure[perturbation size $c=1$]{\includegraphics[width=0.24\textwidth]{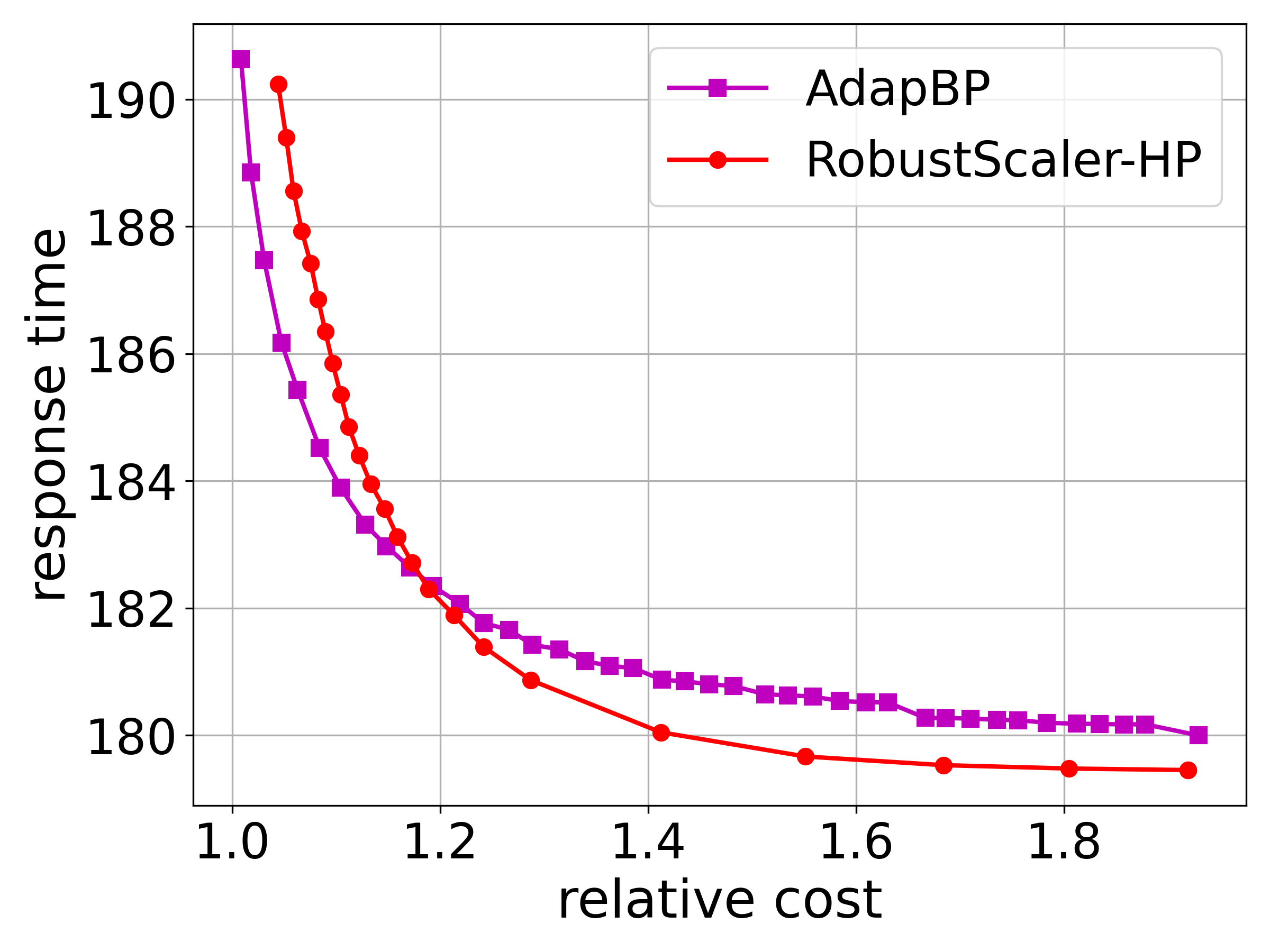}\label{fig:perturb 1 rt}}
    \subfigure[perturbation size $c=2$]{\includegraphics[width=0.24\textwidth]{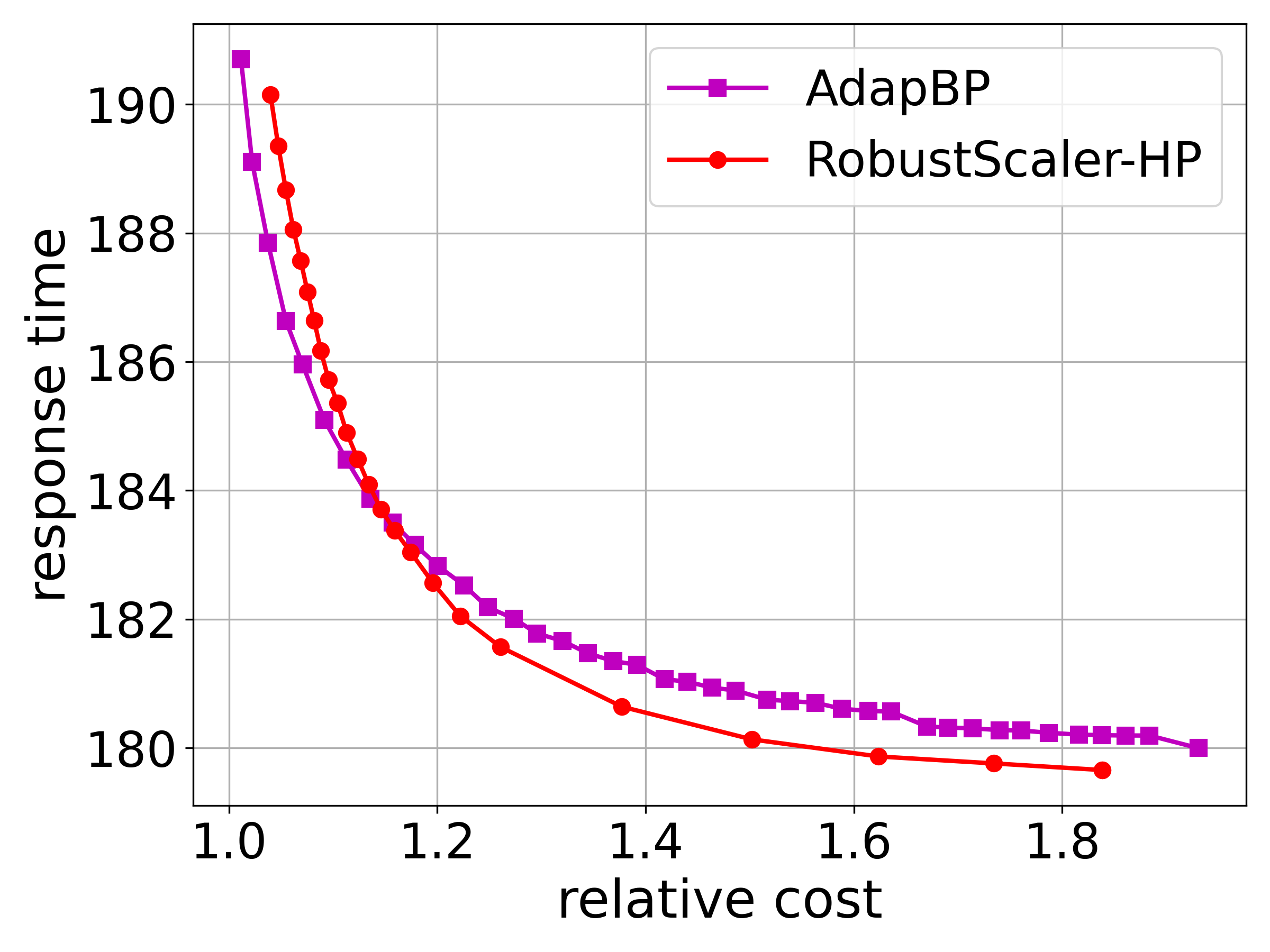}\label{fig:perturb 2 rt}}\\\vspace{-0.1cm}
    \subfigure[perturbation size $c=4$]{\includegraphics[width=0.24\textwidth]{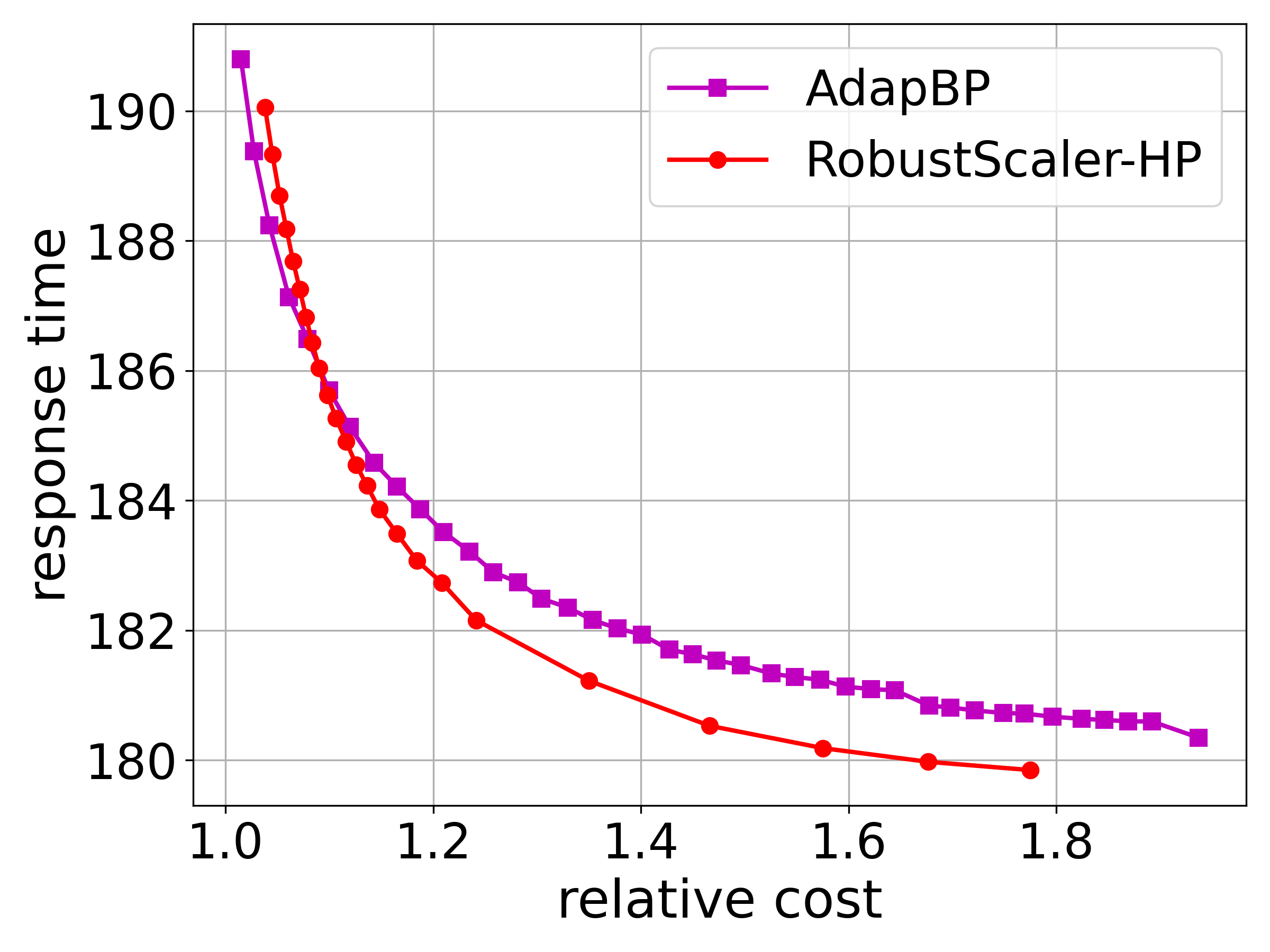}\label{fig:perturb 4 rt}}
    \subfigure[perturbation size $c=6$]{\includegraphics[width=0.24\textwidth]{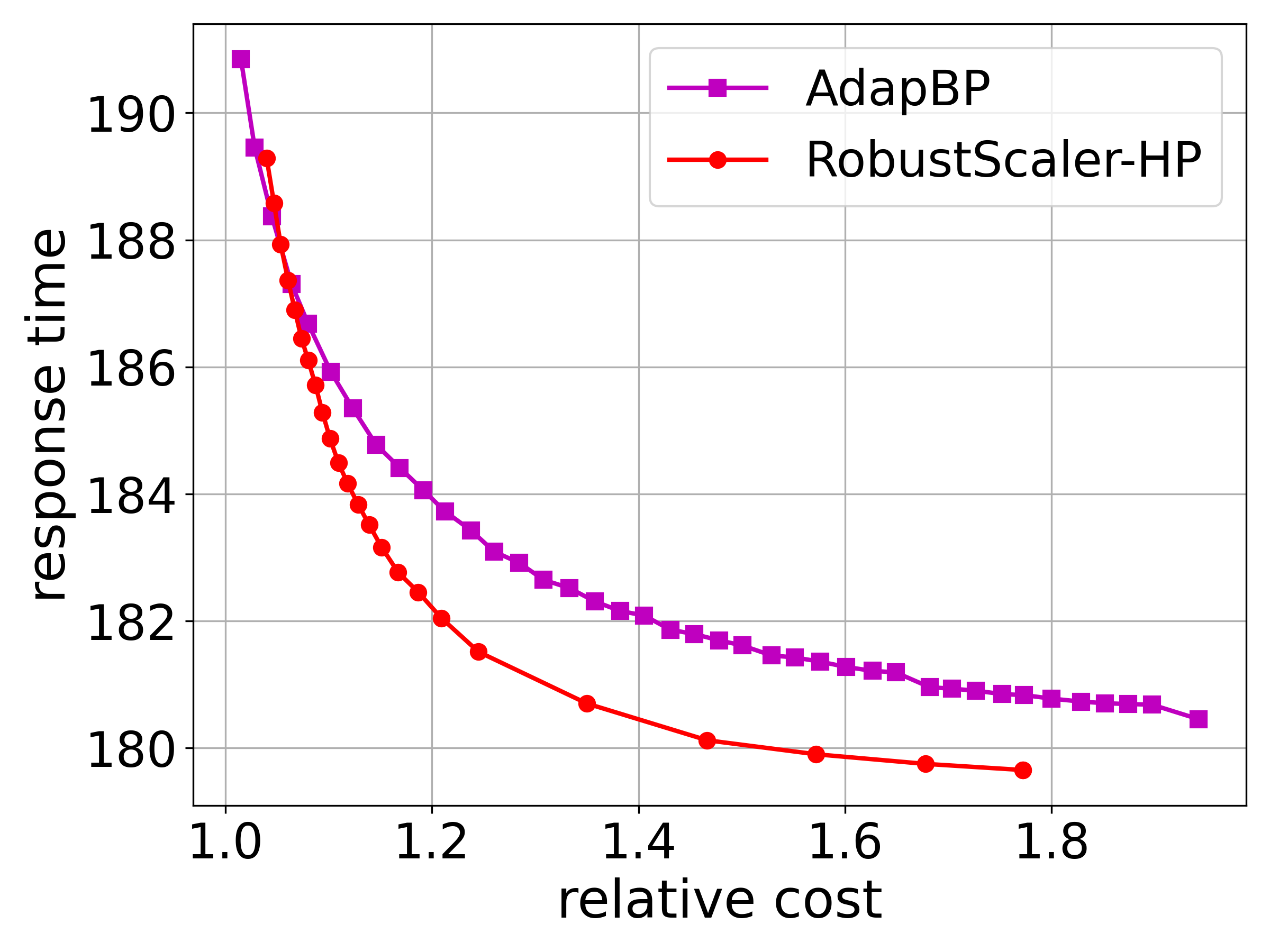}\label{fig:perturb 6 rt}}
    \vspace{-4mm}
    \caption{Comparison of rt\_avg and relative\_cost under perturbed data.}
    \label{fig:perturb_plot_rt}
\end{figure}

\begin{figure}[!t]
    \centering
    \subfigure[perturbation size $c=1$]{\includegraphics[width=0.24\textwidth]{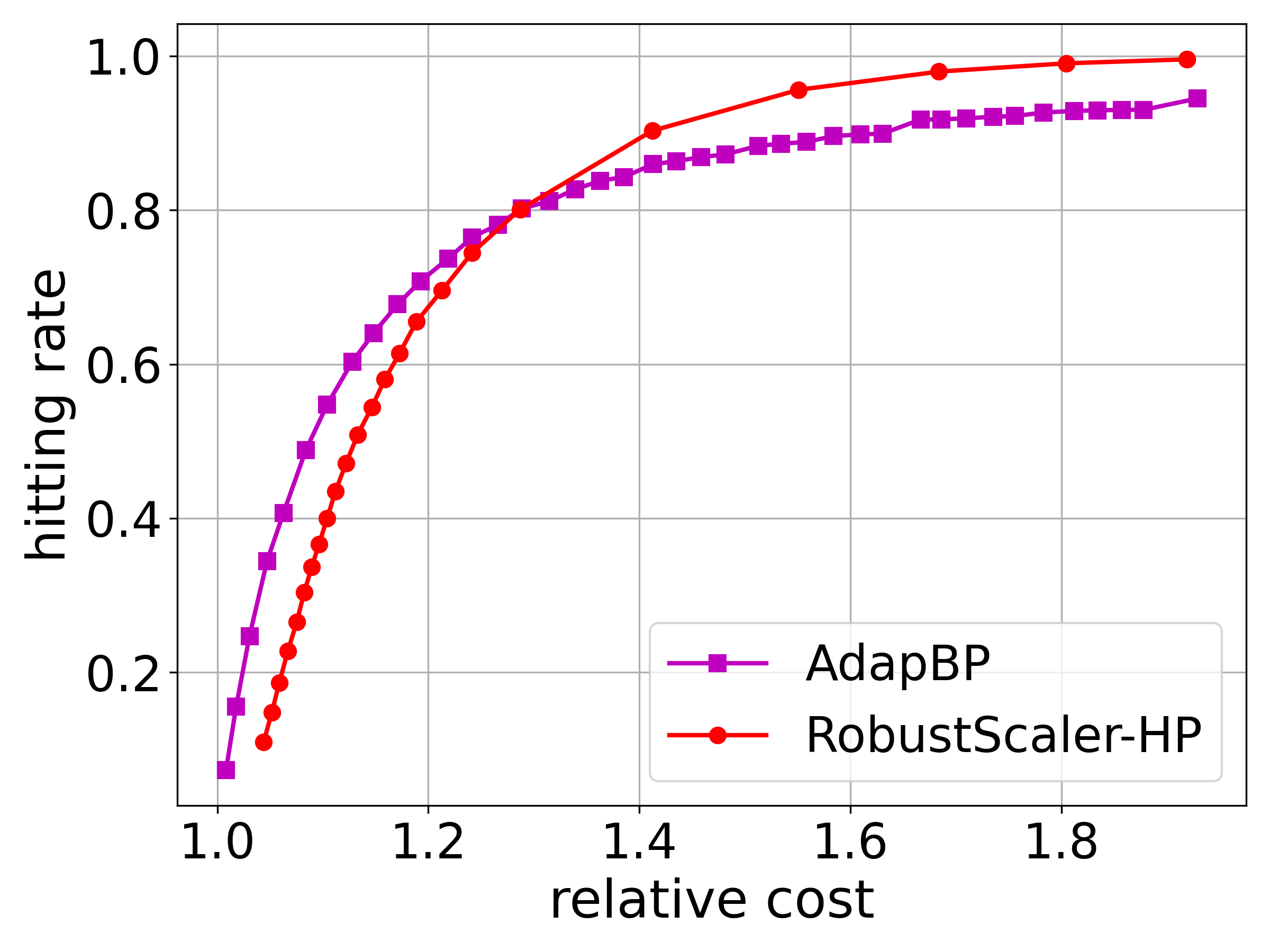}\label{fig:perturb 1 hp}}
    \subfigure[perturbation size $c=2$]{\includegraphics[width=0.24\textwidth]{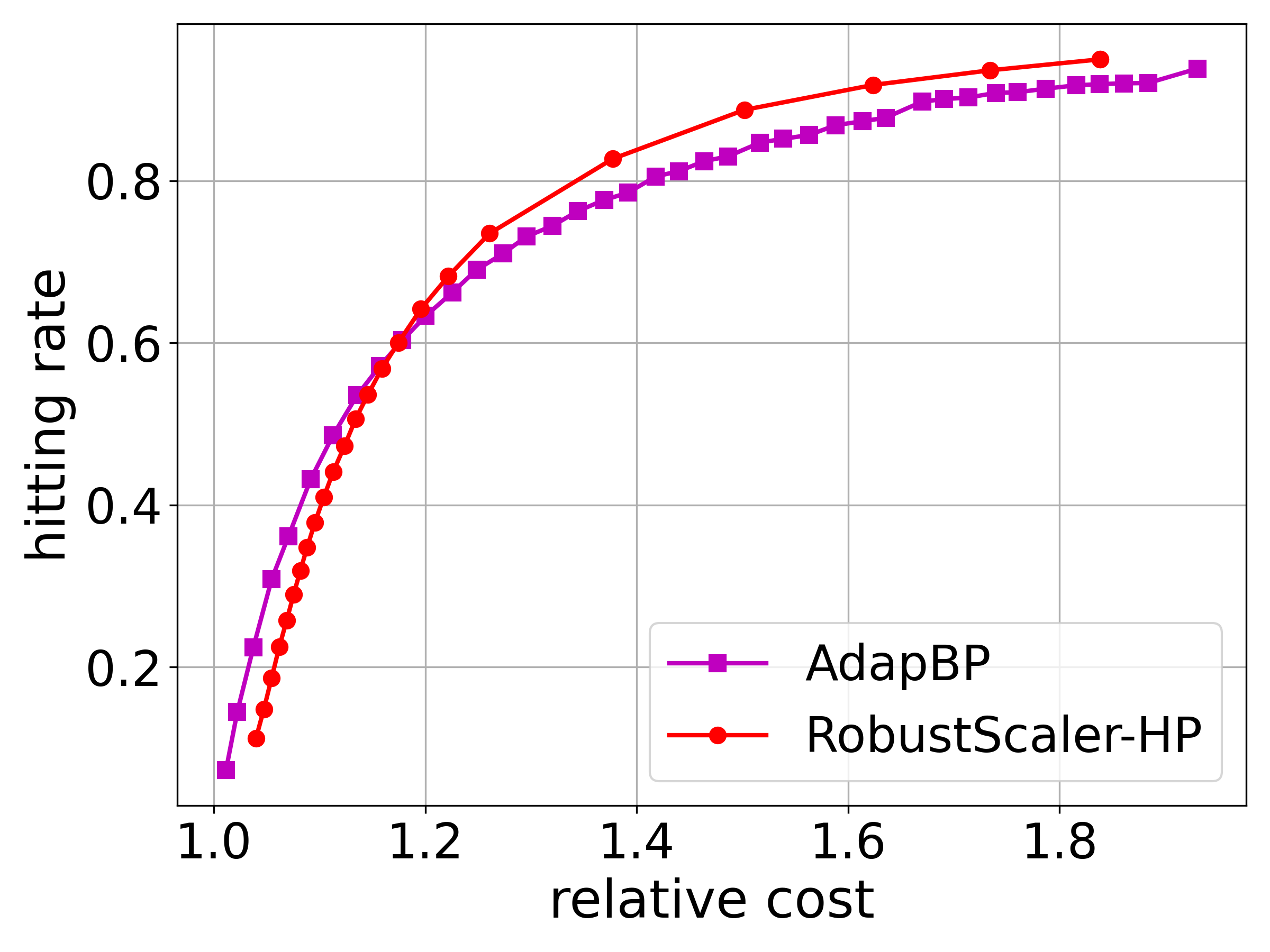}\label{fig:perturb 2 hp}}\\\vspace{-0.1cm}
    \subfigure[perturbation size $c=4$]{\includegraphics[width=0.24\textwidth]{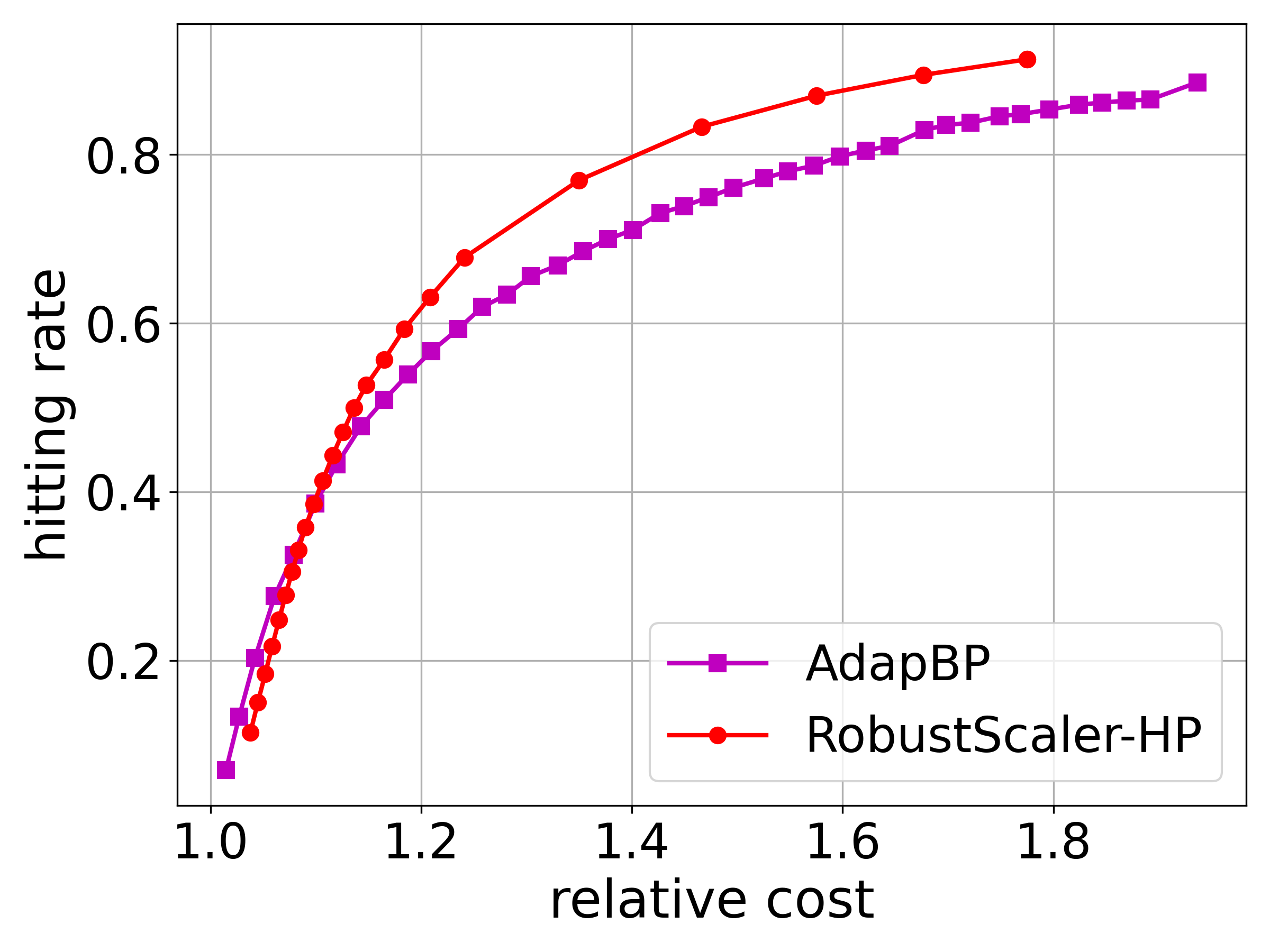}\label{fig:perturb 4 hp}}
    \subfigure[perturbation size $c=6$]{\includegraphics[width=0.24\textwidth]{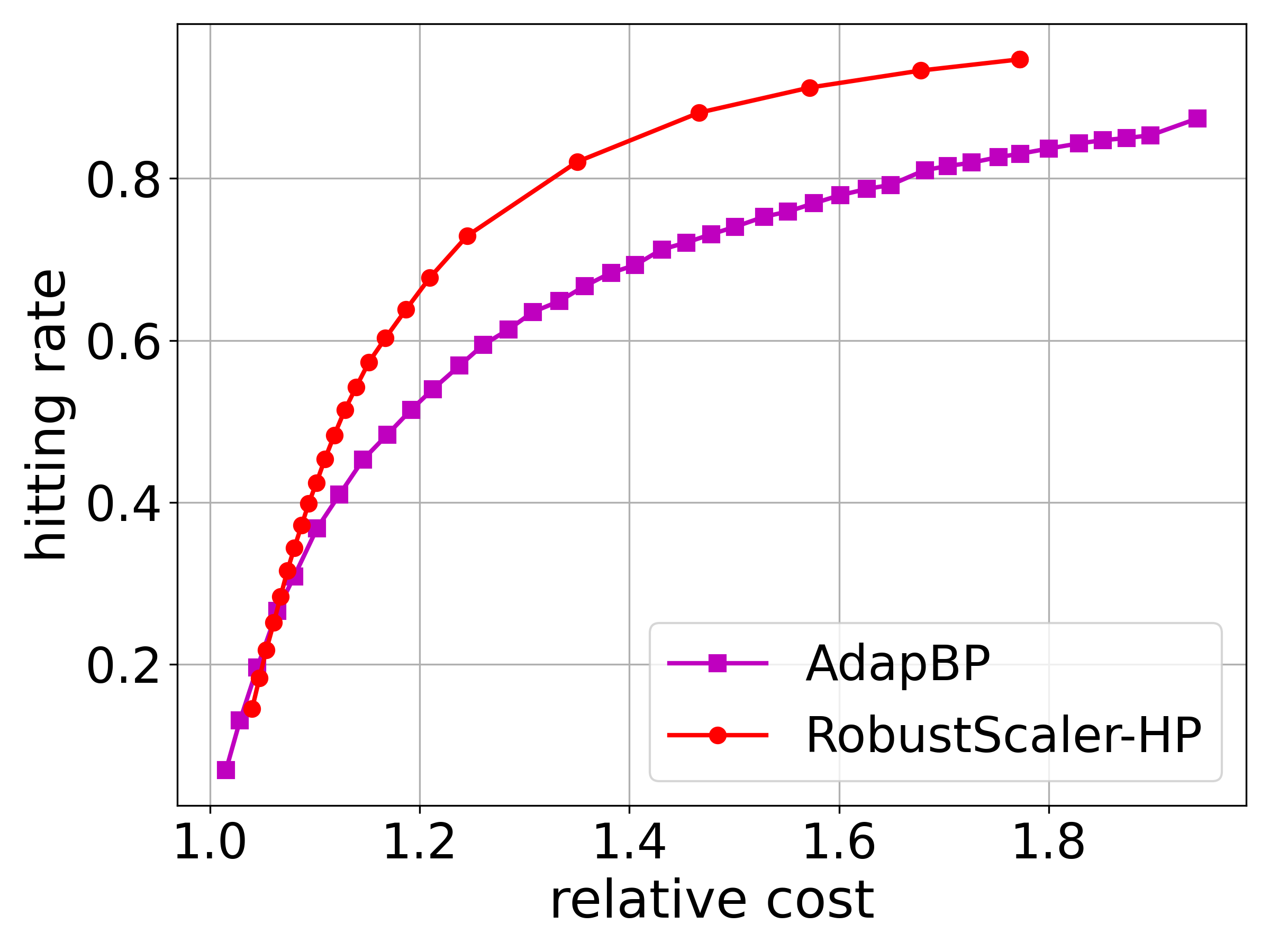}\label{fig:perturb 6 hp}}
    \vspace{-4mm}
    \caption{Comparison of hit\_rate and relative\_cost under perturbed data.}
    \label{fig:perturb_plot_hp}
\end{figure}

\vspace{-1mm}
\subsection{Experimental Results}\vspace{-1mm}
\subsubsection{Comparisons of Autoscaling Methods}
A main goal of the experiments is to compare the Pareto efficiency of these autosclers in hanlding the QoS-cost trade-off. To this end, for each of these autoscalers we vary the parameter that controls this trade-off over a sufficiently wide range and analyze the corresponding QoS and cost metrics. Specifically, in BP the pool size $B$ is varied from $0$ to $8$ for CRS trace, from $0$ to $450$ for Alibaba trace, and from $0$ to $40$ for the Google trace, so that the resulting hitting probability covers a sufficiently wide range in $[0,1]$. In AdapBP the pre-fixed constant is varied from zero to hundreds for each trace. Similarly, in our three RobustScaler variants the corresponding constraint values (the target HP, RT, or cost) are varied. The planning frequency $\Delta$ is set to $1$ second for our RobustScalers.

To visualize the comparison, two types of Pareto plots, hit\_rate vs relative\_cost and rt\_avg vs relative\_cost, are given for each dataset. On each Pareto plot, each line represents a certain autoscaler and each point on the line is the (hit\_rate, relative\_cost) or (rt\_avg, relative\_cost) pair generated by the autoscaler with a particular value of the controlled parameter. Therefore, in the hit\_rate vs relative\_cost plot, the closer a line stays to the top-left, the more efficient the corresponding autoscaler is in the sense that a higher hitting probability is achieved with the same cost. Similarly, in the rt\_avg vs relative\_cost plot lines that are closer to the bottom-left corner are better autoscalers.


Figure \ref{fig:Pareto_plot} summarizes all Pareto plots, in which Figures \ref{fig:pareto of hit and cost on acr}, \ref{fig:pareto of hit and cost on alibaba}, \ref{fig:pareto of hit and cost on google} are hit\_rate vs relative\_cost plots and Figures \ref{fig:pareto of rt and cost on acr}, \ref{fig:pareto of rt and cost on alibaba}, \ref{fig:pareto of rt and cost on google} are rt\_avg vs relative\_cost plots. These plots clearly shows the QoS-cost trade-off for each autoscaler. E.g., in CRS trace, as the pool size of $BP$ increases from $0$ to $8$, the relative cost increases to $2.2$ whereas the hit\_rate increases from $0$ to $1$ and rt\_avg goes down from above $192$ to below $180$. Similar phenomena persist for our RobustScaler variants as the target hitting probability, response time, or cost varies, as well as for AdapBP.

Compared to the heuristic strategy BP, RobustScaler-HP and RobustScaler-RT are consistently better in that they achieve higher hitting rate or lower average RT than BP under the same cost in all the considered cases.
All these gains are tied to our stochastically constrained optimization formulation that explicitly quantifies and optimizes the trade-off. RobustScaler-cost also outperforms BP in most cases, except for some cases with large relative costs on the CRS trace. However, we argue that in these high-cost scenarios the potential marginal gain in hitting probability and response time is already quite small, e.g., the hit\_rate is already around $95\%$ and the rt\_avg only slightly above $180$ (minimum is slightly below $180$) when the lines of RobustScaler-cost and BP cross. Therefore from a user's perspective a relatively low overhead cost might be preferred because of the larger marginal gain in QoS, for which all our RobustScaler variants are more efficient than BP. AdapBP outperforms BP on the CRS and Alibaba traces, which is expected since the pool size is regularly adjusted to match the QPS to avoid wasting instances when queries are rare, however slightly underperforms BP on the Google trace.

Compared to AdapBP, all our RobustScaler variants are superior on the Alibaba and Google traces as they all achieve better response time (Figures \ref{fig:pareto of rt and cost on alibaba}, \ref{fig:pareto of rt and cost on google}) or hitting probability (Figures \ref{fig:pareto of hit and cost on alibaba}, \ref{fig:pareto of hit and cost on google}) under the same cost. On the CRS trace (Figures \ref{fig:pareto of hit and cost on acr}, \ref{fig:pareto of rt and cost on acr}) the result is mixed: The RobustScaler variants underperform AdapBP when the cost is relatively low, but gradually catch up with or even surpass AdapBP (e.g., RobustScaler-HP and RobustScaler-RT) as the cost increases. Although RobustScalers do not perform the best in low-cost cases, they deliver a much stabler hitting probabilities and response times across time as the plots of QoS variance in Figure \ref{fig:var_plot} show. Specifically, Figure \ref{fig:var of rt on acr} is plotted as follows: We first collect the response times of all the queries, and with the queries ordered by their times of arrival we average the response times of every $50$ queries, and then calculate the variance of all these averaged response times which is plotted against the mean response time of all queries. Figure \ref{fig:var of hit on acr} is constructed via a similar procedure on hitting rate. Therefore, each line there clearly shows the variability of QoS at different mean QoS levels as each method is applied with different parameter values. It can be observed that RobustScaler-HP and RobustScaler-RT have much smaller variances hence stabler QoS than AdapBP, and RobustScaler-cost lies in between.

Motivated by the unstable QoS of AdapBP, we conduct further experiments to compare AdapBP and RobustScaler as perturbations of growing sizes are introduced into the CRS trace, and the results summarized in Figures \ref{fig:perturb_plot_rt} and \ref{fig:perturb_plot_hp} show that AdapBP is more sensitive to data changes and its performance quickly deteriorates to a worse level than RobustScalers as the perturbation size grows. The CRS trace is perturbed as follows: On one hand, starting from the beginning of the trace, for every one hour queries within a five-minute time window are deleted; on the other hand, starting from the sixth minute of the trace, for every one hour $c$ more times of queries are added to a five-minute time window. AdapBP and RobustScaler-HP are then applied to the perturbed trace with increasing perturbation size $c=1,2,4,6$. Their relative performance in terms of response time and hitting rate are plotted in Figures \ref{fig:perturb_plot_rt} and \ref{fig:perturb_plot_hp}, respectively, both of which clearly show that as $c$ increases from $1$ to $6$ RobustScaler-HP is closing the performance gap at low-cost scenarios and finally becomes superior to AdapBP globally. This demonstrates the weakness of the simple heuristic AdapBP in handling complex workloads and the robustness of our RobustScaler.

Regarding the choice between the three RobustScaler variants, we find that none of them dominates the other two in all scenarios, but we do see that RobustScaler-HP and RobustScaler-RT perform very similarly on all the three traces. Figure \ref{fig:var_plot} also shows that RobustScaler-HP and RobustScaler-RT deliver stabler QoS than RobustScaler-cost, which is related to the imposed constraints on QoS metrics in their mathematical formulations \eqref{opt:service-constrained} and \eqref{opt:service-constrained using RT}. Therefore, we generally recommend RobustScaler-HP and RobustScaler-RT, whereas RobustScaler-cost is prefered in scenarios with a strict cost budget constraint since RobustScaler-cost places an explicit constraint on the cost (see Section \ref{sec:constraint accuracy} and Figure \ref{fig:cost match} for the accuracy of cost control).


\begin{figure}[!t]
    \centering
    \includegraphics[width=0.35\textwidth]{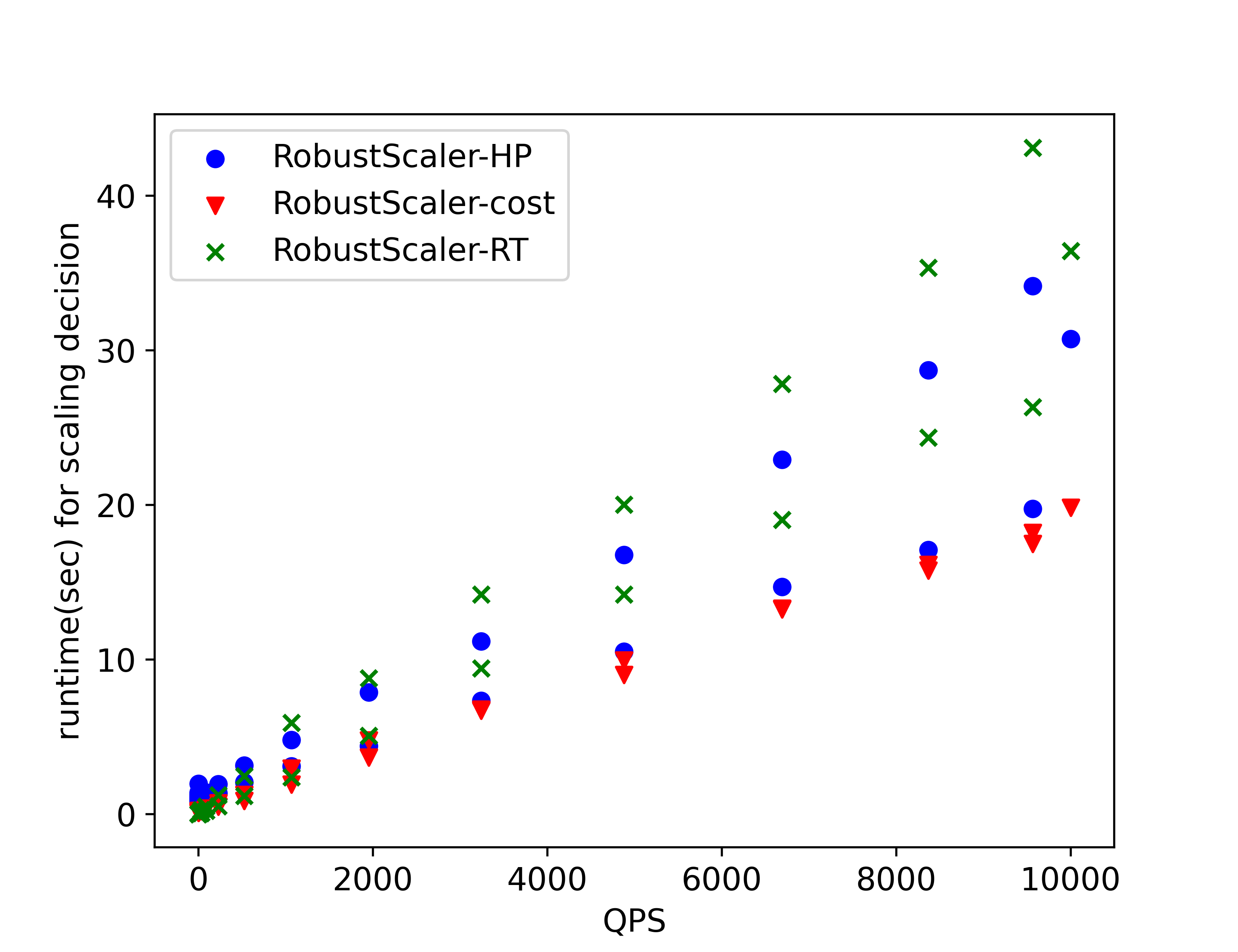}\vspace{-3mm}
    \caption{Runtime (sec) of solving \eqref{sol:service-constrained}\eqref{sol:service-constrained using RT}\eqref{sol:cost-constrained} for scaling decisions versus QPS.}
    \label{fig:scalability comparison}
\end{figure}
\subsubsection{Scalability Towards High QPS}
We investigate how favorably the computation cost of our framework scales with the system workload. In the three real-world traces with a typical QPS between $0.03$ and $6.3$, the training time of modules 1-3 of our model is around $100$ seconds with three weeks of data (CRS trace) and no more than $7$ seconds with four days of data (Alibaba trace), which is short enough given that the NHPP model only needs to be retrained at a low frequency (e.g., every half an hour), and the runtime of the fourth module to generate scaling decisions is under $5$ millisecond for all the three traces which therefore allows for almost instant updates of scaling decisions to accommodate rapidly changing traffic.


In order to further test the computational efficiency when facing even higher QPS, we expand the experiment to a simulated trace with a much higher QPS level up to $10000$. Specifically, we use intensity function $\lambda(t)=1000\cdot 4^{40}(\frac{t\mod 3600}{3600})^{40}(1- \frac{t\mod 3600}{3600})^{40} + 0.001$ for $t\in [0,25200]$ that has an exact period of $3600$ seconds to generate query arrival data over a duration of $7$ hours. Note that the intensity, or the average QPS, ranges from $0.001$ to $10000$ which is way beyond that of the three real-world trace. A fixed pod pending time of $13$ seconds is used, and an exponentially distributed query processing time with a mean of $20$ seconds is used in simulation. The first $6$ hours of the generated data are then used as training data and the last $1$ hour as test data to evaluate our framework. All the three variants (RobustScaler-HP, RobustScaler-RT, RobustScaler-cost) are tested, where  scaling decisions are updated every $5$ seconds using Monte Carlo approximation with a fixed sample size $1000$. The runtime of each decision update is summarized in Figure \ref{fig:scalability comparison}, and the corresponding QoS metrics are shown in Table \ref{tab:performance of constraint satisfaction on simulated data.}.

The scattered plot for runtime and QPS in Figure \ref{fig:scalability comparison} clearly demonstrates the high scalability of all the three variants of RobustScaler. In particular, the method can still deliver scaling decision in seconds even when the QPS is in thousands which is far beyond that of the three real-world traces. The plot also shows a linear growth of the runtime relative to QPS, which is in accordance with our theoretical investigations in Section \ref{sec:opt formulations}. In the case of higher QPS, the computation of decisions takes longer and hence the decisions are delayed and may become obsolete by the time they become available. This, however, can be alleviated by extending the planning horizon to compute instance creations for the next ``$\Delta + \text{delay}$'' instead of $\Delta$ seconds (see the description in Section \ref{sec:autoscaling methods intro}) to compensate for the delay, or by using less Monte Carlo samples in Algorithm \ref{algo:monte carlo} to speed up the computation. Overall the computational burden of RobustScalers scales favorably with the workload size. In addition, such a level of scalability does not come at the expense of the accuracy of scaling decisions, since the Monte Carlo approximated decisions with sample size $1000$ seem good enough for maintaining a QoS level that is close to the target level set for each variant, as shown in Table \ref{tab:performance of constraint satisfaction on simulated data.}. Note that in Table \ref{tab:performance of constraint satisfaction on simulated data.} the target level, $1$ second, for response time by RobustScaler-RT is set with the processing time excluded, i.e., for the quantity $d-\mu_s$ in \eqref{sol:service-constrained using RT}. Similarly, the target cost level, $2$ seconds per instance, set by RobustScaler-cost, is for the average idling time of an instance.

\begin{table}[!t]
\caption{Accuracy of RobustScalers with Monte Carlo approximation on simulated data}
\vspace{-2mm}
    \centering
    \begin{tabular}{c|c|c|c}
    \hline 
     \!QoS levels\!& \!\!RobustScaler-HP\!\! & \!\!RobustScaler-RT\!\!& \!\!RobustScaler-cost\!\! \\\hline\hline
        \!Target level\! & $0.9$ & $1$ & $2$\\ \hline 
        \!Achieved level\! & $0.99$ & $0.51$ & $2.50$\\ \hline
    \end{tabular} \vspace{-1mm}
    \label{tab:performance of constraint satisfaction on simulated data.}
\end{table}

\subsubsection{Robustness Against Missing Data and Anomalies}
\begin{figure}[!t]
    \centering
    \subfigure[hit\_rate vs relative\_cost on CRS trace ]{\includegraphics[width=0.24\textwidth]{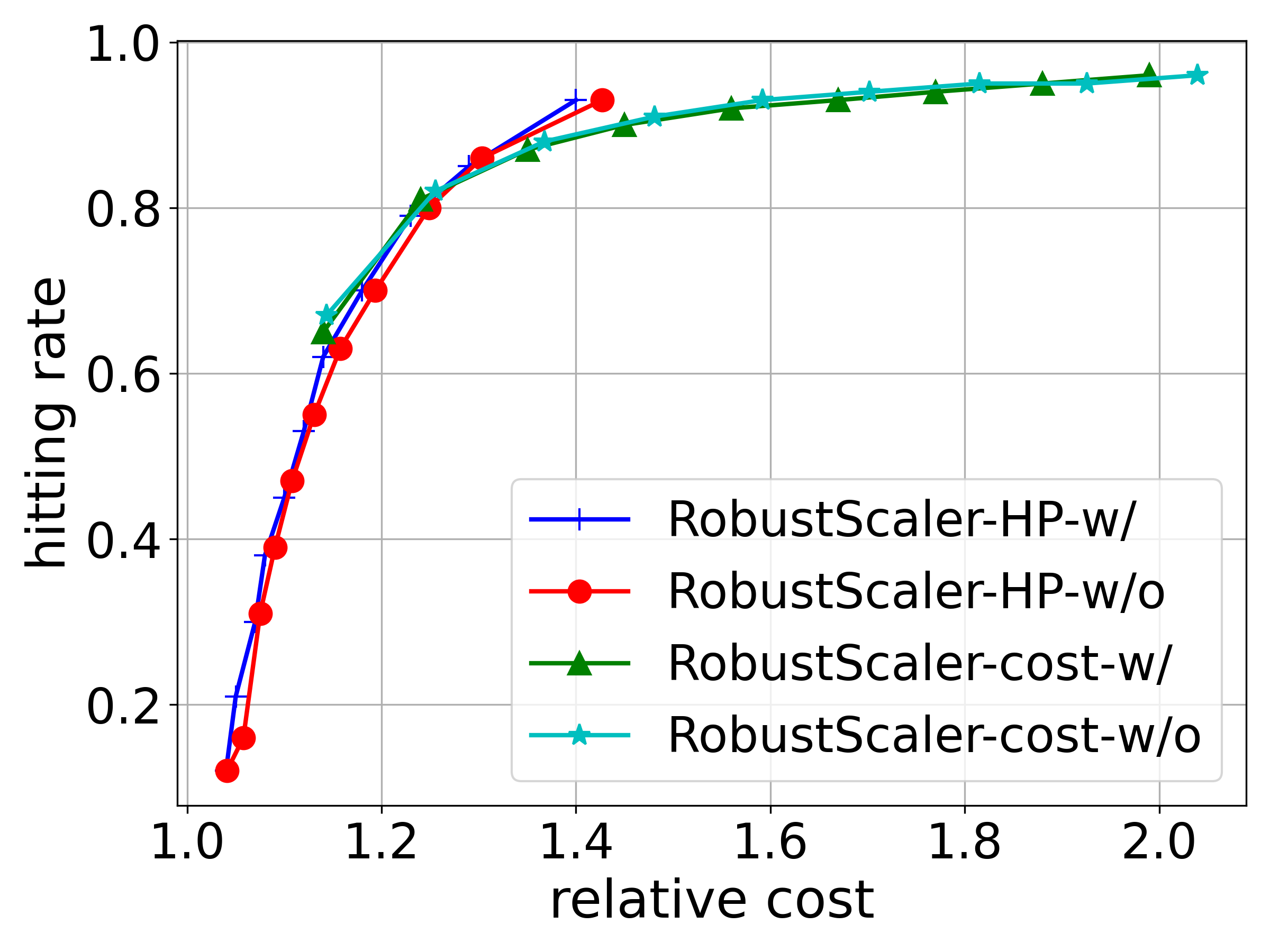}\label{fig:robustness pareto of hit and cost on acr}}
    \subfigure[rt\_avg vs relative\_cost on CRS trace]{\includegraphics[width=0.24\textwidth]{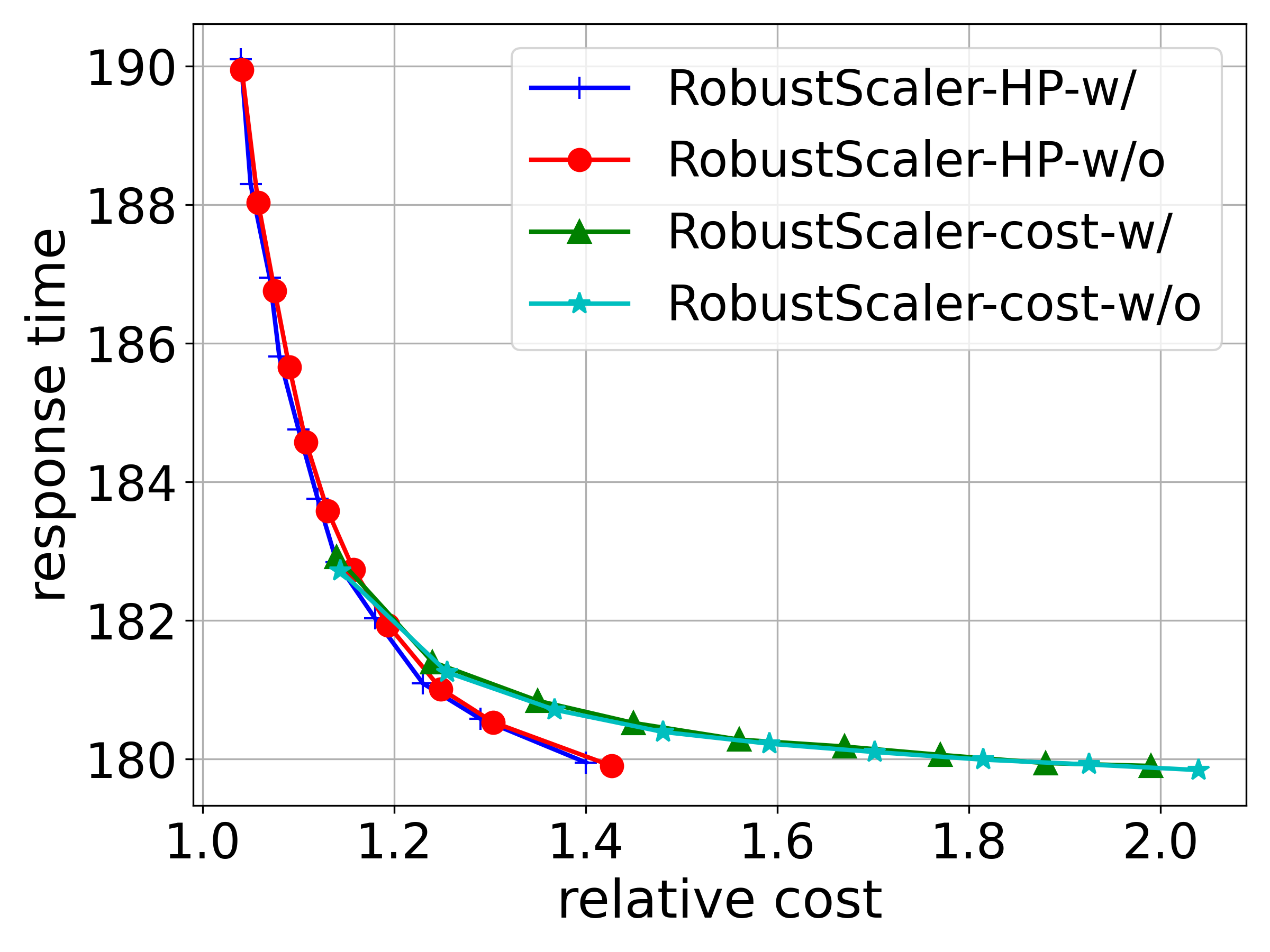}\label{fig:robustness pareto of rt and cost on acr}}\\\vspace{-0.1cm}
    \subfigure[hit\_rate vs relative\_cost on Alibaba trace]{\includegraphics[width=0.24\textwidth]{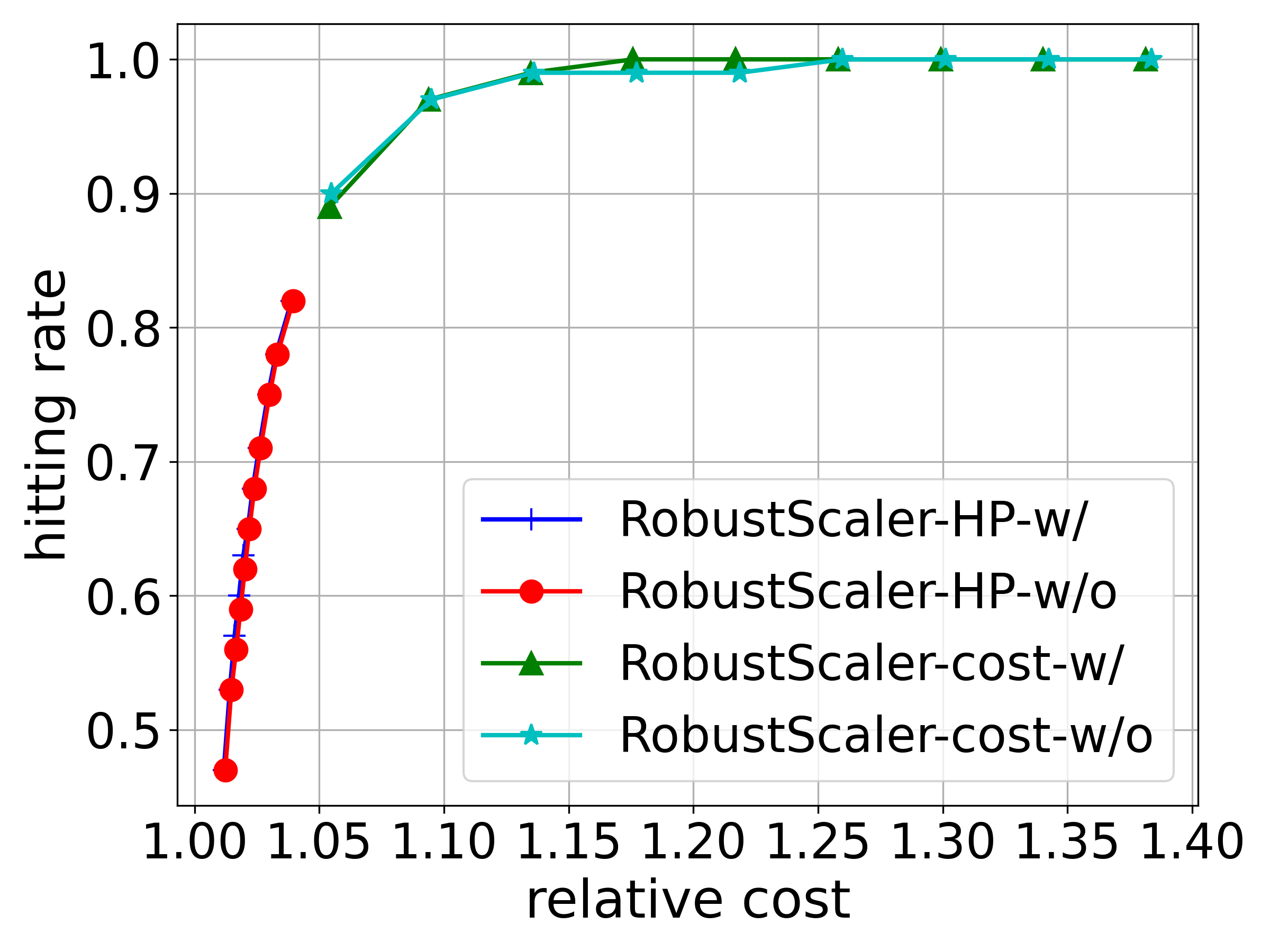}\label{fig:robustness pareto of hit and cost on alibaba}} 
    \subfigure[rt\_avg vs relative\_cost on Alibaba trace]{\includegraphics[width=0.24\textwidth] {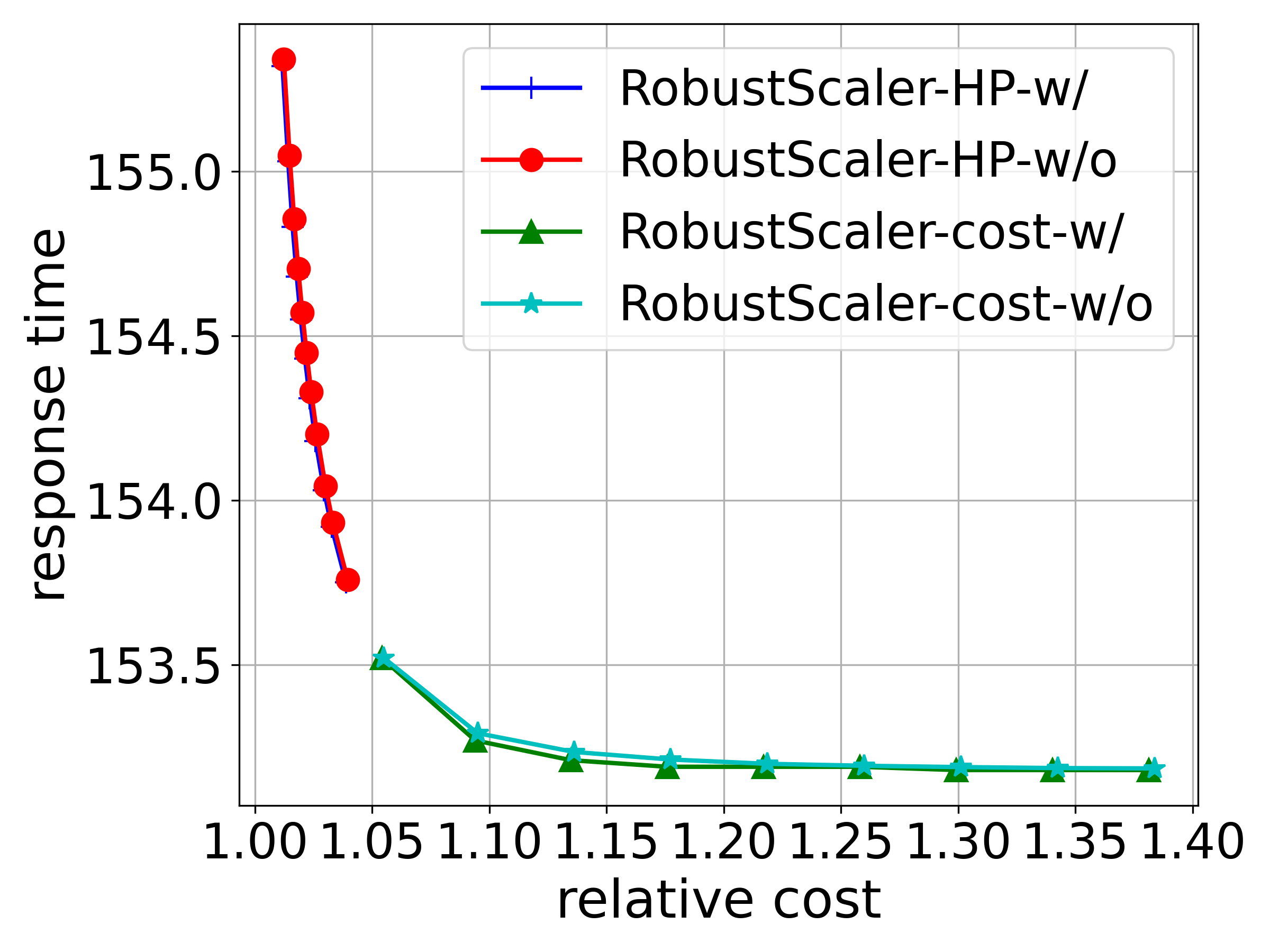}\label{fig:robustness pareto of rt and cost on alibaba}}
    \vspace{-4mm}
    \caption{Comparison before and after anomaly removal or missing data injection, line legends that end with ``w/'' are results with anomaly/missing data, and those ending with ``w/o'' are results without anomaly/missing data.}
    \vspace{-0mm}
    \label{fig:robustness comparison}
\end{figure}

We test the robustness of our autoscalers on Alibaba trace that has an unexpected burst/anomalies, and also on the CRS trace by injecting missing data. In particular, we erase the burst in the Alibaba trace to make the pattern more clear and obvious, and for CRS we inject missing data by removing all the queries in one entire day of the fourth week. We then rerun the experiments with the new modified traces. If the QoS and cost metrics generated by the autoscaler before and after the modifications are similar then the autoscaler can be considered robust against anomalies and missing data. Figure \ref{fig:robustness comparison} summarizes the results for RobustScaler-HP and RobustScaler-cost and it is clear that the resulting QoS and cost metrics are almost identical on the original and modified traces. We also examine the corresponding high-level quantiles ($75\%,95\%,99\% \text{ and } 99.9\%$) of the response times to better characterize the sensitivity to anomalies and missing data. It turns out that both RobustScaler variants have the exactly same response time quantiles at all considered levels before and after removing the anomalies on Alibaba trace, and the changes of the quantiles on the CRS trace are very tiny as shown in Table \ref{tab:quantiles of rt under anomalies.}. In addition, the comparison with AdapBP under perturbed QPS data in Figures \ref{fig:perturb_plot_rt} and \ref{fig:perturb_plot_hp} also reveals the stable performance of RobustScaler under data perturbations. All these demonstrate the robustness of our autoscalers against missing data and anomalies.

\begin{table}[!t]
\caption{Comparison of response time quantiles before and after missing data injection on CRS trace}
\vspace{-2mm}
    \centering
    \begin{tabular}{c|c|c|c|c}
    \hline 
     \multirow{2}{*}{\tabincell{c}{Quantile\\ level}}& \multicolumn{2}{c|}{\!\!RobustScaler-HP\!\!} & \multicolumn{2}{c}{\!\!RobustScaler-cost\!\!} \\\cline{2-5}
     &\!\!w/ missing\!\!&\!\!w/o missing\!\!&\!\!w/ missing\!\!&\!\!w/o missing\!\!\\\hline\hline
        \!$75\%$\! & $182.0$& $182.0$& $181.0$ & $181.0$\\ \hline 
        \!$95\%$\! & $624.0$& $624.0$& $622.6$ & $622.6$\\ \hline
        \!$99\%$\! & $1513.5$& $1512.5$& $1509.5$ & $1507.6$\\ \hline
        \!$99.9\%$\! & $6822.6$& $6822.6$& $6822.6$ & $6822.6$\\ \hline
    \end{tabular} \vspace{-1mm}
    \label{tab:quantiles of rt under anomalies.}
\end{table}

\subsubsection{Accurate Control of QoS/Cost Levels and Effects of Planning Frequency}\label{sec:constraint accuracy}
\begin{figure}[!t]
    \centering
    \subfigure[hitting probability on CRS trace ]{\includegraphics[width=0.24\textwidth]{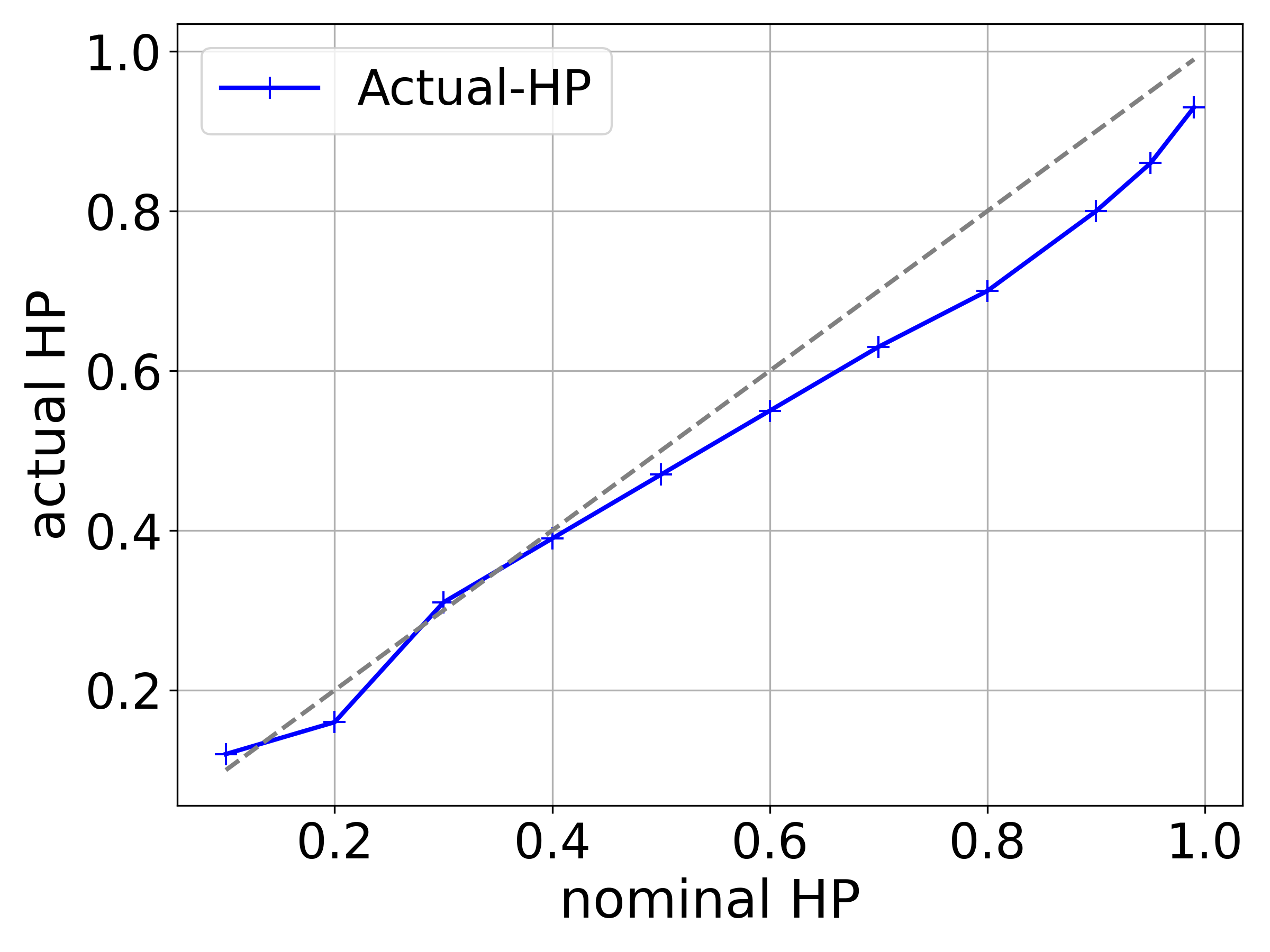}\label{fig:HP match}}
    \subfigure[response time on CRS trace]{\includegraphics[width=0.24\textwidth]{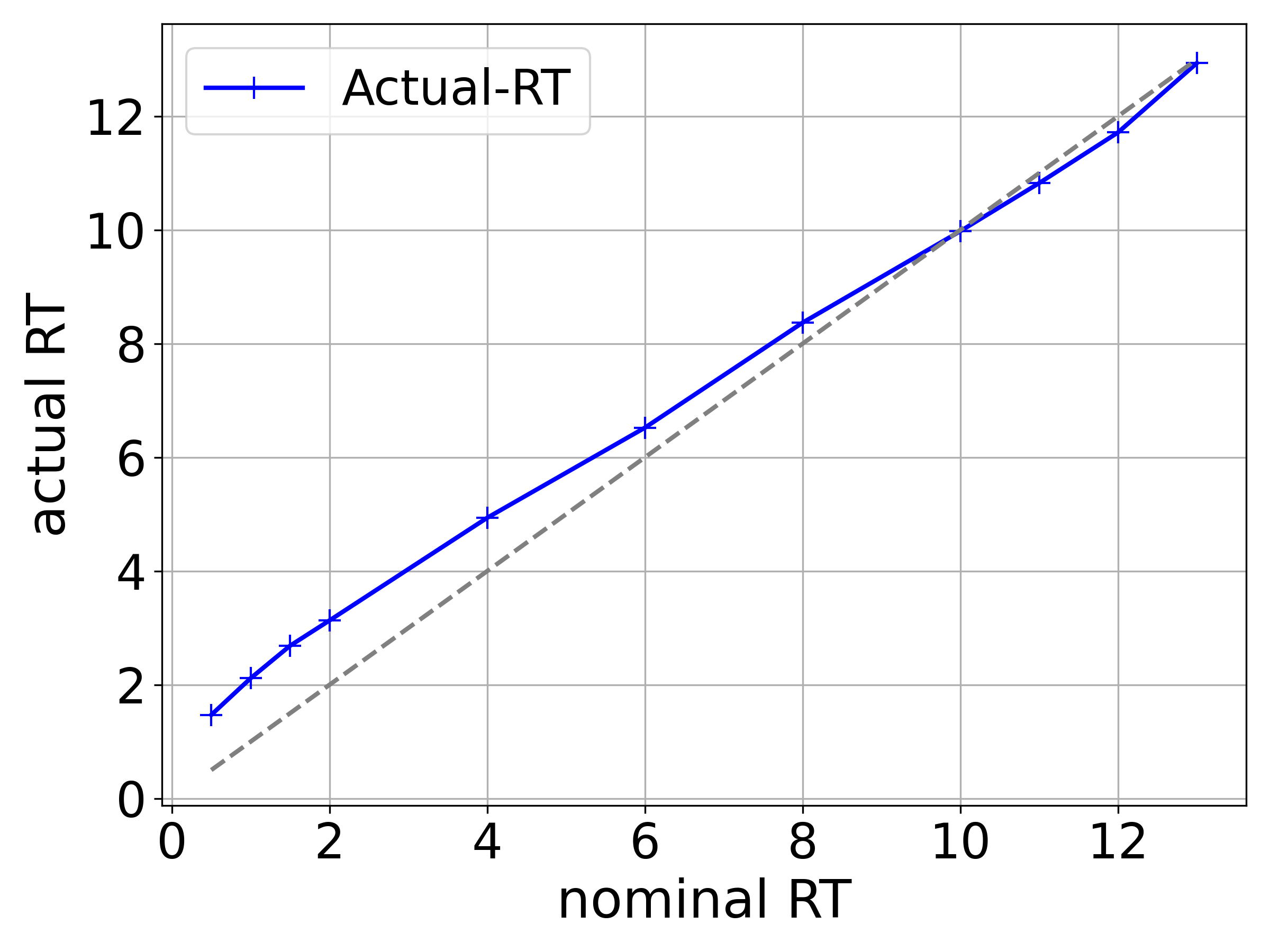}\label{fig:RT match}} \\\vspace{-0.1cm}
    \subfigure[cost on CRS trace]{\includegraphics[width=0.24\textwidth]{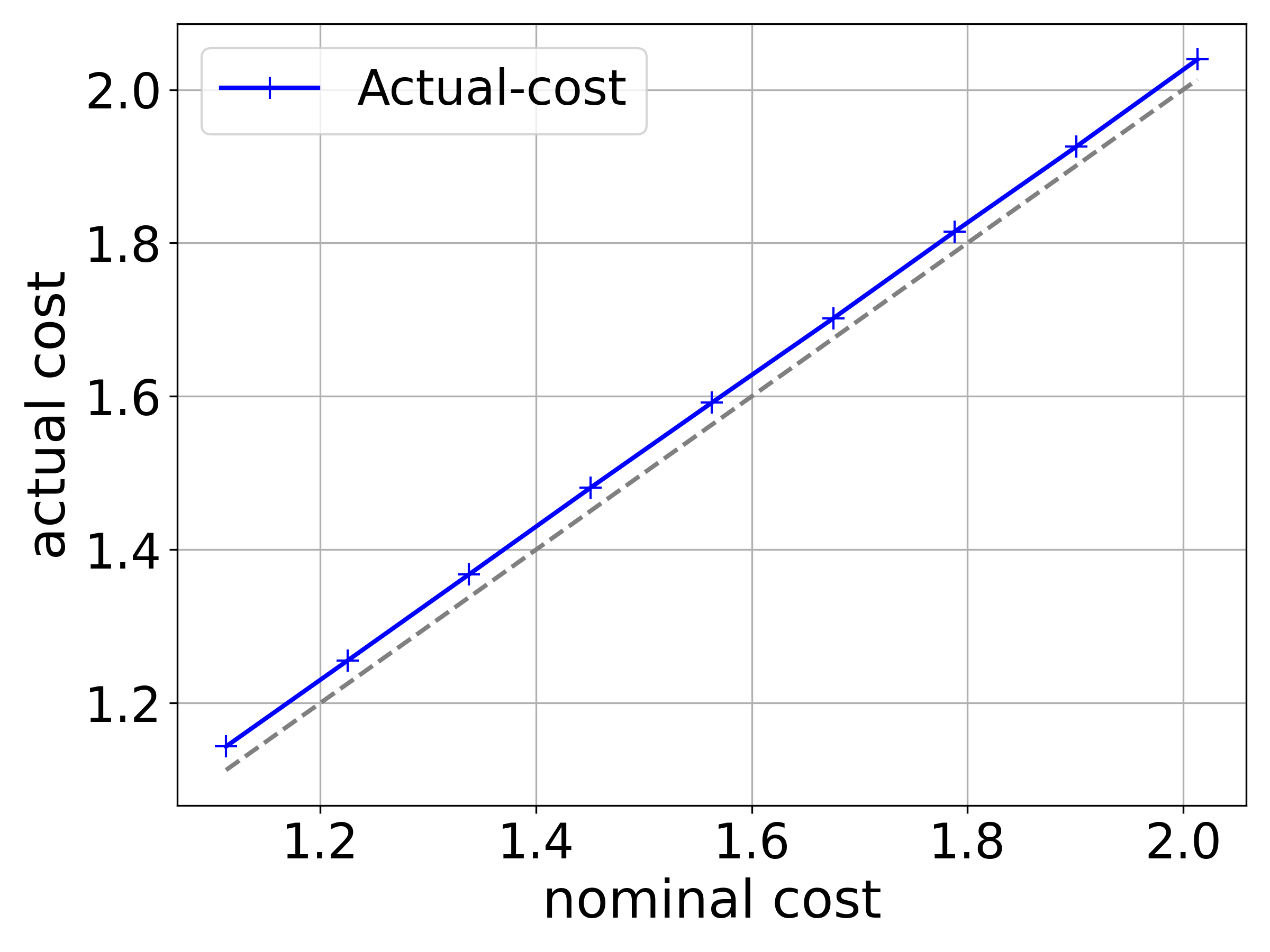}\label{fig:cost match}}
    \subfigure[planning frequency on CRS trace]{\includegraphics[width=0.24\textwidth]{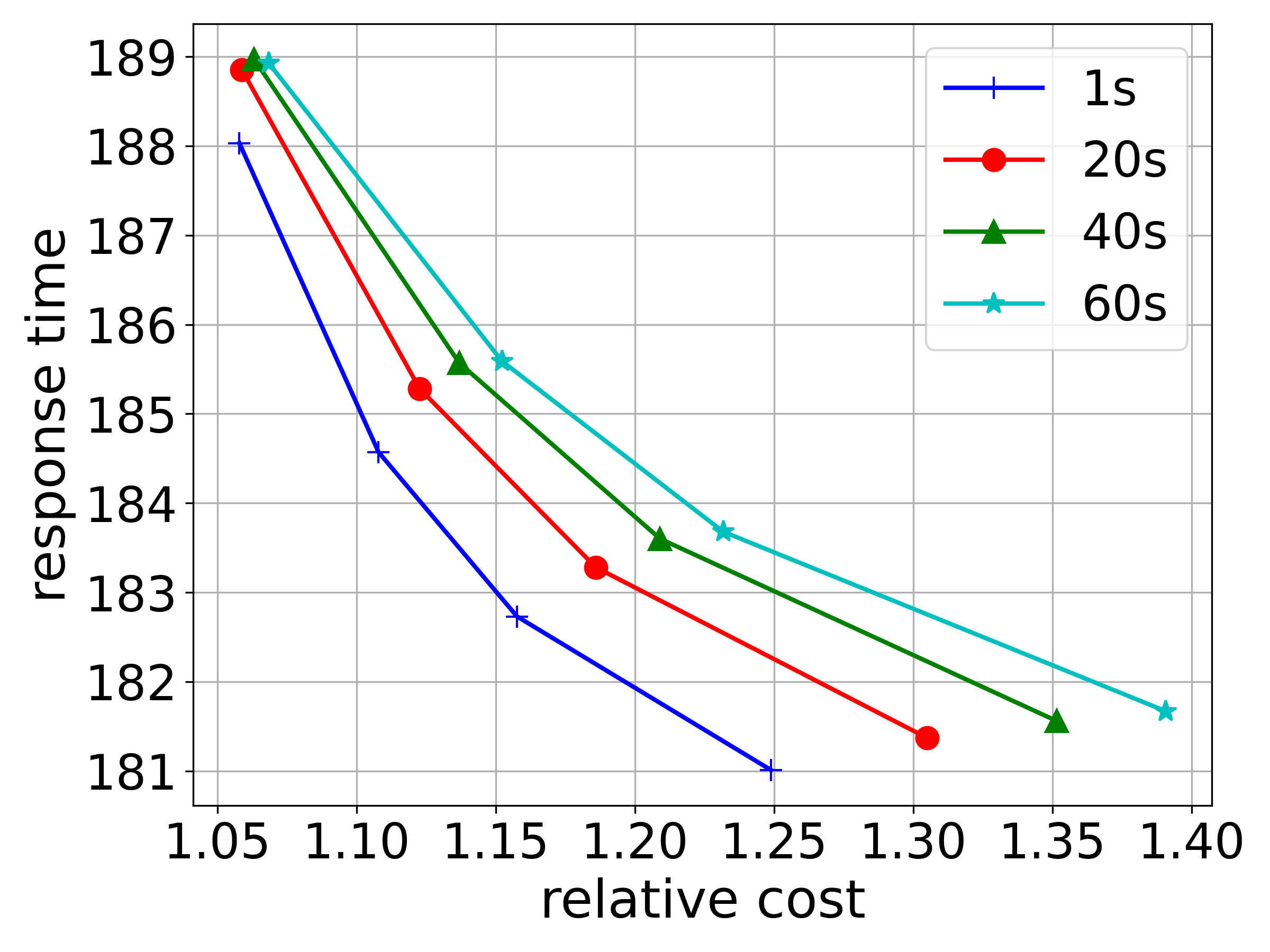}\label{fig:frequency}}
    \vspace{-4mm}
    \caption{Nominal versus actual QoS/cost levels, and efficiency under different planning frequencies.}
    \label{fig:QoS and cost match}
\end{figure}

Our theory suggests that RobustScaler is guaranteed to achieve a pre-specified hitting probability if the arrival process is approximately Poisson, and here we test the accuracy of RobustScaler in maintaining a pre-specified level of QoS in practice on the CRS trace. Figures \ref{fig:HP match}, \ref{fig:RT match}, \ref{fig:cost match} show the comparison of the actual HP, RT, and cost values collected from the results and the corresponding nominal values used in our RobustScalers. Ideally, these two sets of values shall be on the straight line $y=x$ which is marked as dotted line in each plot, and we see that they are indeed very close to the line, demonstrating the advantage of our RobustScalers in achieving a promised level of QoS/cost. 
Another aspect of our autoscalers that we study here is the planning frequency. We mentioned that more frequent decision planning leads to more savings in cost, and to validate this claim on real data, we increase the $\Delta$ of RobustScaler-HP from $1$ to $60$ seconds and the corresponding results are summarized in Figure \ref{fig:frequency}. It clearly shows that as the planning interval $\Delta$ gets larger, the more costly the decisions will be in order to attain the same level of response time.

Apart from the stochastically constrained optimization used to generate scaling decisions, another key ingredient that affects the accuracy of scaling decisions is the estimation accuracy of the query arrival intensity, as explained by our Proposition \ref{prop:HP error}. To quantitatively examine the impact of our periodicity regularization on the estimation accuracy, we compare the mean squared error (MSE) and mean absolute error (MAE) of the intensity estimate given by \eqref{NHPP formulation w/ periodicity} and another estimate by the same loss but without the periodicity regularization term. Specifically, we generate arrival data from the ground truth intensity $\lambda(t)=4^{10}(\frac{t\mod 86400}{86400})^{10}(1-\frac{t\mod 86400}{86400})^{10}+0.1$ for $t\in[0,604800]$ with a period length of $86400$, and apply \eqref{NHPP formulation w/ periodicity} with or without the periodicity regularization to train the intensity model. The errors of the intensity estimates are summarized in Table \ref{tab:accuracy comparison w/ and w/o reg.}, which shows that the periodicity regularization boosts the accuracy for $40$-$60\%$ as measured by MSE or MAE. This demonstrates the advantage of the periodicity regularization in practical scenarios where periodicity is common.

\begin{table}[!t]
\caption{Impact of periodicity regularization on NHPP intensity error}
\vspace{-2mm}
    \centering
    \begin{tabular}{c|c|c|c}
    \hline 
    Metric    & NHPP w/o reg. & NHPP w/ reg. & improvement \\\hline\hline
        MSE & $5.08\times10^{-4}$ & $2.24\times 10^{-4}$ & $56\%$\\ \hline 
        MAE & $1.53\times 10^{-2}$ & $9.30\times 10^{-3}$ & $39\%$\\ \hline
    \end{tabular} \vspace{-1mm}
    \label{tab:accuracy comparison w/ and w/o reg.}
\end{table}

\subsubsection{Deployment in A Real Environment}
We also deploy and test our framework in a real-world autoscaled system. In the simulated environment, the time elapsed in computing scaling decisions are not taken into account in the autoscaling procedure and hence the obtained decisions never become obsolete, e.g., a scaling decision that says ``create a pod in $5$ seconds from now'' is assumed available and executed exactly at the prescribed time even if the computation of the decision itself takes longer than $5$ seconds in physical time. Therefore a key difference of real environments from the simulated one, among others, is that not only the scaling decision itself but also the runtime of our algorithm execution will directly affect the scaling process.


The CRS trace is used to generate queries from a client and then send to a server. When the server receives a query, it finds an available instance if there is any or create a new one in an Alibaba Serverless Kubernetes (ASK) cluster, and the instance then sleeps for a certain amount of time to mimic the processing time of a real meaningful query. When the sleeping ends, the instance is immediately deleted. Each instance is built with the Alpine container image with a 4-core CPU and 8GB RAM. We test the performance of RobustScaler-HP in autoscaling the system with a target hitting probability level of $0.9$. Like in the previous simulated environment, the first three weeks of the trace are used as training data and the last week as testing.
\begin{table}[!t]
\caption{Comparison of RobustScaler-HP in simulated and real environments}
\vspace{-2mm}
    \centering
    \begin{tabular}{c|c|c|c}
    \hline 
    Environment    & HP & RT & cost \\\hline\hline
        Simulated & $0.80$ & $181.0$ & $240.3$\\ \hline 
        Real & $0.83$ & $189.3$ & $228.7$\\ \hline
    \end{tabular}\vspace{-1mm} 
    \label{tab:comparison simulatetd vs real.}
\end{table}
Table \ref{tab:comparison simulatetd vs real.} summarizes the performance of RobustScaler-HP under real and simulated environments. It can be seen that the achieved hitting probabilities, response times, and costs in seconds are close to each other. This shows that our algorithm continues to work as expected in real environments, and that thanks to the high computational efficiency of our method the adversarial impact of delay of scaling decisions is minimal.





\vspace{-1mm}
\section{CONCLUSION}\label{sec:conclusion}
\vspace{-2mm}

In this paper we propose a novel proactive autoscaling scheme RobustScaler for the scaling-per-query cloud computing scenario, which achieves superior trade-off between cost and QoS and is robust to noise, missing data and anomalies.
We leverage NHPP with specialized regularization techniques to flexibly capture both periodicity and stochasticity of query arrivals with an efficient ADMM solution. Furthermore, we formulate a stochastically constrained optimization and a sequential scaling scheme with provable probabilistic performance guarantees to achieve better scaling decisions than heuristic strategies. Our extensive experiments demonstrate the effectiveness and efficiency of RobustScaler in both simulated and real environments.



\bibliographystyle{IEEEtran}
\bibliography{main}

\end{document}